\documentclass[letterpaper,12pt]{article}
\usepackage[utf8]{inputenc} 
\usepackage{amsmath}
      \allowdisplaybreaks 
      \everymath{\displaystyle} 
\usepackage{mathtools} %
\usepackage{amsfonts}
\usepackage{mathrsfs}
\usepackage{amssymb}
\usepackage{multicol} 
\usepackage{geometry}
\usepackage{graphicx} 
\usepackage{float} 
\usepackage{subcaption} 
\usepackage{verbatim} 
\usepackage{hyperref} 
\usepackage{cancel} 
\usepackage{authblk} 
\usepackage{lscape} 
\usepackage[all,cmtip]{xy} 
\usepackage{float} 
\usepackage[toc,page]{appendix} 

\usepackage{tikz}
\usepackage{tikzscale}
\usetikzlibrary{positioning}
\usetikzlibrary{arrows}
\usetikzlibrary{intersections, backgrounds} 
\usetikzlibrary{calc}

\usepackage{array,longtable}
      \newcolumntype{C}{>{$}c<{$}}  
      \newcolumntype{R}{>{$}r<{$}}  
      \newcolumntype{L}{>{$}l<{$}}  
\usepackage[titles]{tocloft} 
\usepackage[nottoc,notlot,notlof]{tocbibind} 

\usepackage{titlesec}
\titleformat{\chapter}[display]
{\normalfont\huge\bfseries}{\chaptertitlename\ \thechapter}{20pt}{\Huge}   
\titlespacing*{\chapter}{0pt}{-15pt}{15pt} 

\makeatletter
\let\@fnsymbol\@alph
\makeatother

\newcommand\numberthis{\addtocounter{equation}{1}\tag{\theequation}} 

\title{Star-triangle type relations from $2d$ $\mathcal{N}=(0,2)$ $USp(2N)$ dualities}
\author[]{\normalsize J. de-la-Cruz-Moreno\thanks{E-mail: jdlcruz@fis.cinvestav.mx}}
\author[]{\normalsize H. García-Compeán\thanks{E-mail: compean@fis.cinvestav.mx}}
\affil[]{\it \normalsize Physics department, Centro de Investigación y de Estudios Avanzados del Instituto Politécnico Nacional, P. O. Box 14-740, C. P. 07000, Mexico city, Mexico}

\date{}
\begin{document}
\maketitle
      \begin{abstract}

Inspired by the gauge/YBE correspondence this paper derives some star-triangle type relations from dualities in $2d$ $\mathcal{N}=(0,2)$ $USp(2N)$ supersymmetric quiver gauge theories. To be precise, we study two cases. The first case is the Intriligator-Pouliot duality in $2d$ $\mathcal{N}=(0,2)$ $USp(2N)$ theories. The description is performed explicitly for $N=1,2,3,4,5$ and also for $N=3k+2$, which generalizes the situation in $N=2,5$. For $N=1$ a triangle identity is obtained. 
For $N=2,5$ it is found that the realization of duality implies slight variations of a star-triangle relation type (STR type). The values $N=3,4$ are associated to a similar version of the asymmetric STR.
The second case is a new duality for $2d$ $\mathcal{N}=(0,2)$ $USp(2N)$ theories with matter in the antisymmetric tensor representation that arises from dimensional reduction of $4d$ $\mathcal{N}=1$ $USp(2N)$ Csáki-Skiba-Schmaltz duality. It is shown that this duality is associated to a triangle type identity for any value of $N$. In all cases Boltzmann weights as well as interaction and normalization factors are completely determined. Finally, our relations are compared with those previously reported in the literature.

      \end{abstract}

\newpage
\tableofcontents

\newpage
\section{Introduction}

Classical and quantum integrable systems defined through the Yang-Baxter equation (YBE) have been studied from diverse view points in the literature. Important work has been summarized in many compendia reported at early stages, see for instance, \cite{Baxter:1982zz,Jimbo:1983nw,Jimbo:1989mc,Yang:1989kj,Yang:1994qs,Gomez:1996az}. 

Recently, a surprising relation between quiver gauge theories in various dimensions with diverse degrees of supersymmetry and integrable models in statistical mechanics has starting to be explored by many authors, for an overview see \cite{Yamazaki:2018xbx} and references therein. This relation is termed in the literature as the {\it gauge/YBE correspondence}. In this correspondence the underlying spin lattice in the integrable model is identified to the quiver diagram of the quiver gauge theory. Moreover the self-interaction and nearest-neighbour interaction of spins correspond to the gauge vector supermultiplets in the adjoint representation of the gauge group and the chiral multiplet in the bifundamental representation of the gauge group, respectively. As a result of the work on this subject a dictionary of this correspondence has been established between the structure and features of the integrable models and the quiver field theory. For instance, the spin variables can be identified with the gauge holonomies along non-trivial homology $1$-cycles, the rapidity line can be identified with the zig-zag path, the spectral parameter with the R-charge, the statistical partition function to the field theory partition function, the star-star relation to the Seiberg(-like) duality, the Yang-Baxter equation with the Yang-Baxter duality, etc.

There are plenty of integrable models that have been obtained from supersymmetric dualities via the gauge/YBE correspondence \cite{Yamazaki:2018xbx}. This have been done for different dimensions, amounts of supersymmetry, gauge groups and diverse curved manifolds. To state some examples, there are integrable models associated to $2d$ $\mathcal{N}=(2,2)$ theories (see, for instance, \cite{Yamazaki:2015voa,Jafarzade:2017fsc}), $3d$ $\mathcal{N}=2$ theories (see, for instance, \cite{Eren:2019ibl,Gahramanov:2016ilb}) and $4d$ $\mathcal{N}=1$ theories (see, for instance, \cite{Spiridonov:2010em,Yamazaki:2013nra}). A large list of more dualities is given in \cite{Gahramanov:2017ysd}.
Despite the rich zoo of new integrable models obtained from the gauge/YBE correspondence and as far as we know, there are no explicit integrable models associated with $2d$ $\mathcal{N}=(0,2)$ theories and we have found very few literature about this topic (see, for example, \cite{Yagi:2015lha} for the context of brane constructions).

The above considerations have motivated the authors to study $2d$ supersymmetric $\mathcal{N}=(0,2)$ field theories. It would be interesting to investigate whether this family of theories can be incorporated to the context of the mentioned correspondence and to check if there are integrable models that can be associated to these models. However, as a first step in this direction we will concentrate  in the present work in studying what kind of star-triangle relations can be associated to some of the dualities obeyed for these supersymmetric theories. Thus the aim of the present article is to study what kind of star-triangle relations (or some of their variants as the star-tringle type relation, STR type, or the triangle identity) arises from some dualities in supersymmetric quiver gauge theories. In this direction, we first analyse the Intriligator-Pouliot duality in $2d$ $\mathcal{N}=(0,2)$ $USp(2N)$ theories coming from dimensional reduction (see \cite{Gadde:2015wta} for description of this reduction) of $4d$ $\mathcal{N}=1$ $USp(2N)$ confining Intriligator-Pouliot theory originally studied in \cite{Intriligator:1995ne}. The analysis is carried out for different values of $N$. We also study a new duality for $2d$ $\mathcal{N}=(0,2)$ $USp(2N)$ theories with matter in the antisymmetric tensor representation found in \cite{Sacchi:2020pet} that arises from a dimensional reduction of $4d$ $\mathcal{N}=1$ $USp(2N)$ Csáki-Skiba-Schmaltz duality first discussed in \cite{Csaki:1996eu}.
The expressions obtained in this work share many features with the standard star-triangle relation, such as the general distribution and dependence on the spin variables and spectral parameters, although they have not exactly the same form. In certain cases we found some similarity with star-triangle type relations discussed in the literature of Yang–Baxter/$3D$-consistency correspondence \cite{Bazhanov:2016ajm,Kels:2019ktt,Kels:2018xge,Kels:2020zjn}.

Intriligator-Pouliot and Csáki-Skiba-Schmaltz dualities are important in the context of Seiberg-like duality \cite{Spiridonov:2008zr,Spiridonov:2014cxa} while $2d$ $\mathcal{N}=(0,2)$ theories have special interest since the discovery of trialities among them \cite{Gadde:2013lxa,Dedushenko:2017osi}. In a remarkable paper \cite{Gadde:2013lxa} the authors studied the space of $2d$ $\mathcal{N}=(0,2)$ supersymmetric quiver gauge theories and there it was found a {\it triality} among them. There it is also speculated the possibility that the triality would be associated with the tetrahedron equation of statistical mechanics (see \cite{Stroganov:1997br}, for instance, for some work on this equation) in a similar way that Seiberg's duality is related with Yang-Baxter equation.

This article is organized as follows: in Section 2 a brief overview of the gauge/YBE correspondence is given. In Section 3 we review the $4d$ Csáki-Skiba-Schmaltz duality and its associated STR type expression. Section 4 is devoted to obtain some slight variants of the star-tringle type relations from $2d$ theories with supersymmetry ${\cal N}=(0,2)$ and $USp(2N)$ gauge group. For $2d$ Intriligator-Pouliot duality this is carried out for the first five values of $N$ and a general case with $N=3k+2$. In this same section it is shown that the relation associated with $2d$ Csáki-Skiba-Schmaltz duality for any value of $N$ is a triangle type identity. Finally, in Section 5 we give our final remarks. 


\section{Overview of gauge/YBE correspondence}

In the present section we provide a brief overview of the gauge/YBE correspondence. Our aim will not intend to be exhaustive but only to introduce the notation and conventions that will be useful in the subsequent sections. 

As stated in \cite{Yamazaki:2018xbx}, an integrable model is considered to be a solution of the Yang-Baxter equation with spectral parameters that satisfies the rapidity difference property in their $R$-matrices
            \begin{align}
            R_{23}(z_2-z_3) R_{13}(z_1-z_3) R_{12}(z_1-z_2) = R_{12}(z_1-z_2) R_{13}(z_1-z_3) R_{23}(z_2-z_3),
            \label{YBEwithspectralparametersandrapiditydifferenceproperty}
            \end{align}
where $z_1$, $z_2$ and $z_3$ are the spectral parameters and
            \begin{align}
            R_{ij} \in \textrm{End}(V_i \otimes V_j)
            \label{operadorRij}
            \end{align}
for all $i,j \in \{ 1, 2, 3 \}$ with $i \neq j$. Note that in equation (\ref{YBEwithspectralparametersandrapiditydifferenceproperty}) operators (\ref{operadorRij}) are actually promoted to operators in $\textrm{End}(V_1 \otimes V_2 \otimes V_3)$ by an adequate insertion of an identity. 
            
One of the best known integrable models is the $2d$ Ising model in statistical mechanics which is part of the Ising-type integrable models that can be obtained from the YBE depending on the values taken by the spin variables, which can be discrete, continuous or a combination of both of them. There are two relations from statistical mechanics, known as star-star relation and star-triangle relation (SSR and STR from now on, respectively), such that a solution of one of them is immediately a solution of the YBE.
In constructing integrable models it is preferable to solve one of those relations instead because of the highly constrained nature of the YBE.
            
The so called gauge/YBE correspondence is given between supersymmetric quiver gauge theories and integrable models in statistical mechanics. It is then necessary to roughly describe such theories and their relation with statistical mechanics.

\subsection{Construction of the correspondence}

The review of this subsection is carried out mainly following Ref. \cite{Yamazaki:2018xbx}. For a quiver gauge theory in dimension $d$ with gauge group $G$, let $V$ and $E$ be the sets of all vertices and edges in its associated quiver  diagram, respectively. Each vertex $v$ contains gauge fields\footnote{Here, $\mu=1, \dots, d$ and $x$ is a point in the $d$-dimensional space. From now on it will be written as $A_v$.}
$A_v^{\mu}(x)$ with values in the associated Lie algebra of the gauge group $G_v$ while each edge $e$ from $v'$ to $v''$ contains a matter field $\phi_e$ transforming in the bifundamental representation $\left( \Box, \overline{\Box} \right)$ of $G_{v'} \times G_{v''}$. The partition function for the quiver gauge theory with gauge group $G$ is given by the partition function which is the product of partition functions for all possible vertices and edges
            \begin{align}
            \widetilde{\mathcal{Z}} = \int \prod_{v_ \in V} DA_v \prod_{e_ \in E} D\phi_e
            \ e^{- \displaystyle \widetilde{\mathcal{L}} \left( \{ A_v \}_{v \in V}, \{ \phi_e \}_{e \in E} \right) },
            \label{partitionfunctionquivergauge}
            \end{align}
where
            \begin{align}
            \widetilde{\mathcal{L}} \left( \{ A_v \}_{v \in V}, \{ \phi_e \}_{e \in E} \right) =
            \sum_{v \in V} \widetilde{\mathcal{L}}^v \left( A_v \right) + \sum_{e \in E} \widetilde{\mathcal{L}}^e \left( \{ A_v \}_{v \in e}, \phi_e \right),
            \label{lagrangianquivergauge}
            \end{align}
with $\widetilde{\mathcal{L}}^v \left( A_v \right)$ the kinetic term for $A_v$ and $\widetilde{\mathcal{L}}^e \left( \{ A_v \}_{v \in e}, \phi_e \right)$ the interaction term between $\phi_e$ and the gauge fields $\{ A_v \}_{v \in e}$.

To obtain a supersymmetric quiver gauge theory one must supersymmetrize the theory (\ref{partitionfunctionquivergauge}) by using, for instance, supersymmetric localization. After this procedure is applied to  $4d$ $\mathcal{N} = 1$ Yang-Mills theory, each vertex $v$ has now associated a $\mathcal{N} = 1$ vector multiplet\footnote{Each field is in the adjoint representation of $G_v$.}
$\mathcal{V}_v = (A_v, \lambda_v, F_v)$ where $\lambda_v$ and $F_v$ are a gaugino and an auxiliary field, respectively, while each edge $e$ from $v'$ to $v''$ has now a $\mathcal{N} = 1$ 
chiral multiplet\footnote{The multiplet is in a non-trivial representation of $G_{v'} \times G_{v''}$.}
$\Phi_e = (\phi_e, \psi_e, H_e)$ where $\psi_e$ and $H_e$ are a fermion and an auxiliary field, respectively. The supersymmetric theory has partition function
            \begin{align}
            \mathcal{Z} = \int \prod_{v_ \in V} DA_v D\lambda_v DF_v \prod_{e_ \in E} D\phi_e D\psi_e DH_e
            \ e^{- \displaystyle \mathcal{L} \left( \{ \mathcal{V}_v \}_{v \in V}, \{ \Phi_e \}_{e \in E} \right) }.
            \label{partitionfunctionquivergaugesupersymmetric}
            \end{align}
            After regularization by integration in a compact manifold $M$ this partition function can be reduced to
            \begin{align}
            \mathcal{Z} [ M ] = \sum_{ \{ \sigma_v \}_{v \in V} }
            e^{- \displaystyle \mathcal{L} \left( \{ \sigma_v \}_{v \in V} \right) },
            \label{partitionfunctionquivergaugesupersymmetricreducida}
            \end{align}
where
            \begin{align}
            \mathcal{L} \left( \{ \sigma_v \}_{v \in V} \right) =
            \sum_{v \in V} \mathcal{L}^v \left( \sigma_v \right) + \sum_{e \in E} \mathcal{L}^e \left( \{ \sigma_v \}_{v \in e} \right),
            \label{lagrangianquivergaugesupersymmetricreducida}
            \end{align}
with $\{ \sigma_v \}_{v \in V}$ a set of finite-dimensional variables associated with holonomies of $A_v$ along non-trivial homology cycles of $M$. Equations (\ref{partitionfunctionquivergaugesupersymmetricreducida}) and (\ref{lagrangianquivergaugesupersymmetricreducida}) nicely match with statistical mechanics because in a typical statistical lattice the vertices contain spin variables $s_v$ so that the partition function is given by
            \begin{align}
            \mathcal{Z} = \sum_{ \{ s_v \}_{v \in V} }
            e^{- \displaystyle \xi \left( \{ s_v \}_{v \in V} \right) },
            \label{partitionfunctionstatisticalmechanics}
            \end{align}
where
            \begin{align}
            \xi \left( \{ s_v \}_{v \in V} \right) =
            \sum_{v \in V} \xi^v \left( s_v \right) + \sum_{e \in E} \xi^e \left( \{ s_v \}_{v \in e} \right),
            \label{lagrangianstatisticalmechanics}
            \end{align}
with $\xi^v \left( s_v \right)$ the self-interaction term at vertex $v$ and $\xi^e \left( \{ s_v \}_{v \in e} \right)$ the nearest-neighbour interaction of the spins. Comparison of Eqs. (\ref{partitionfunctionquivergaugesupersymmetricreducida}) and (\ref{lagrangianquivergaugesupersymmetricreducida}) with (\ref{partitionfunctionstatisticalmechanics}) and (\ref{lagrangianstatisticalmechanics}), respectively, provides a deep connection between supersymmetric quiver gauge theories and statistical mechanical theories, and this is an important point of the gauge/YBE correspondence. 

Until now the description has made manifest the correspondence between quiver diagram and statistical lattice, supersymmetric quiver gauge partition function and statistical partition function, vector multiplet in the adjoint representation and self-interaction term, chiral multiplet in the bifundamental representation and nearest-neighbour interaction, and holonomies of gauge fields and spin variables;
but the relation between these sets of theories is actually deeper. Dualities in supersymmetric gauge theories play a very important role  \cite{Yamazaki:2018xbx}. In particular, we will see in the following subsection that Seiberg-like duality is related to the star-star relation.

\subsection{From Seiberg-like duality to star-star relation}

Original Seiberg duality \cite{Seiberg:1994pq} is a strong/weak (or S) duality between two $4d$ $\mathcal{N}=1$ gauge theories in the infrared regimen, one with gauge group $SU(N_c)$ and the other one with gauge group $SU(N_f-N_c)$, where $N_c$ and $N_f$ are the number of colors and flavours. For example, for an original theory with $SU(2)$ gauge group and $SU(6)$ flavour group (this means $6$ flavours or chiral multiplets transforming in the fundamental representation of both the gauge and the flavour groups, and the vector multiplets transforming in the adjoint representation of the gauge group), the dual theory is that with gauge group $SU(4)$, $15$ chiral multiplets in the totally antisymmetric tensor representation of the flavour group and without gauge degrees of freedom. There are several generalizations of this duality depending on the dimension of the theory and the amount of supersymmetry.

As stated at the beginning of this section, a solution of the SSR is also a solution of the YBE. This means that the correspondence between Seiberg-like duality and SSR can be used to build and to study integrable models from the point of view of supersymmetric quiver gauge theories. One way for constructing integrable models is to find the correspondence between supersymmetric indices ({\it a.k.a.} elliptic flavoured genera) of Seiberg-like dual theories and then directly compare them with SSR or STR expressions in order to find the associated Boltzmann weights. 


In Ref. \cite{Jafarzade:2017fsc} an integrable model is derived from Seiberg-like duality of $2d$ $\mathcal{N}=(2,2)$ supersymmetric quiver gauge theories on $\mathbb{T}^2$. This model is shown to be a dimensional reduction of $4d$ $\mathcal{N}=1$ supersymmetric quiver gauge theories on $\mathbb{T}^2 \times \mathbb{S}^2$ \cite{Honda:2015yha}.
Gauge theories $2d$ $\mathcal{N}=(2,2)$ are described in \cite{Witten:1993yc,Melnikov:2019tpl} as dimensional reduction of $4d$ $\mathcal{N}=1$ theories. The spectrum of  $(2,2)$ theories in two dimensions consists of two different multiplets, namely, the chiral multiplet with fermions $\psi_+$ and $\psi_-$ of opposite chirality and a complex scalar $\phi$, and the vector multiplet $V$ containing Majorana fermions $\lambda_+$ and $\lambda_-$, a complex scalar $\sigma$ and gauge bosons $\left\{ v_{\alpha} \right\}_{\alpha=0,1}$.
The analysis of the index (flavoured elliptic genus in the NS-NS sector) of $2d$ $\mathcal{N}=(2,2)$ supersymmetric gauge theories is carried out in \cite{Gadde:2013dda}. Moreover, the contribution due to chiral and vector multiplets is given in terms of Jacobi theta functions, $\theta(y;q)$.
The index duality of these $2d$ $\mathcal{N}=(2,2)$ theories is given as follows \cite{Jafarzade:2017fsc}
            \begin{align}
            \frac{1}{2} \bigg( \frac{(q,q)^2_{\infty}}{\theta(y;q)} \bigg) \int \frac{dz}{2 \pi i z}
            \Bigg[ \frac{ \prod^6_{i=1} \Delta (a_i z^{\pm 1}; q, y)}{\Delta (z^{\pm 2}; q, y)} \Bigg]
            =
            \prod_{1 \leq i < j \leq 6} \Delta (a_i a_j; q, y),
            \label{indexNazari}
            \end{align}
where
            \begin{align}
            \Delta (a; q, y) = \frac{\theta(ay;q)}{\theta(a;q)},
            \end{align}
here, the left hand side consists of a theory with gauge group $SU(2)$ and flavour group $SU(6)$ while the right hand one is a theory with only $15$ chiral multiplets. Note that the field content is essentially the same as in the $4d$ $\mathcal{N}=1$ Seiberg duality\footnote{When considering Seiberg-like dualities in different dimensions one usually neglects the superpotential of the theory and just analyses the field content.}. This $2d$ $\mathcal{N}=(2,2)$ duality corresponds in the statistical mechanical side to STR for continuous spin variables \cite{Jafarzade:2017fsc}
            \begin{align}
            \int d\sigma S(\sigma) W_{\eta - \gamma}(\sigma, \sigma_i) W_{\eta - \beta}(\sigma, \sigma_j) 
                                   W_{\eta - \alpha}(\sigma, \sigma_k)
            =
            R(\alpha, \beta, \gamma) W_{\alpha}(\sigma_i, \sigma_j) W_{\beta}(\sigma_i, \sigma_k) W_{\gamma}(\sigma_j, \sigma_k),
            \label{startrianglerelation}
            \end{align}
where $S(\sigma)$ and $R(\alpha, \beta, \gamma)$ stand for the interaction and normalization factors, respectively, while $W_{\alpha}(\sigma_i, \sigma_j)$ are the associated Boltzmann weights.

The next section contains a derivation of the star-triangle type expression discussed in \cite{Spiridonov:2010em} obtained from $4d$ $\mathcal{N}=1$ $USp(2N)$ Csáki-Skiba-Schmaltz duality for supersymmetric quiver gauge theories with matter in the antisymmetric tensor representation first studied in \cite{Csaki:1996eu}, whose index duality is given in terms of standard elliptic gamma functions.

\section{Star-triangle type relation for \texorpdfstring{$4d$ $\mathcal{N}=1$ $USp(2N)$}{} Csáki-Skiba-Schmaltz duality} 
\label{Integrable model for 4d CSS duality}

In \cite{Spiridonov:2010em} the star-triangle type relation associated with $4d$ $\mathcal{N}=1$ $USp(2N)$ duality for theories with matter in the antisymmetric tensor representation is determined. It is convenient to realize the duality for these theories through the following expression
            \begin{align*}
            &
            \frac{(p,p)^n_{\infty} (q,q)^n_{\infty}}{(4\pi)^n n!} \int\limits_{\mathbb{T}^n} 
            \prod_{j=1}^n 
            \left[ \frac{dz_j}{i z_j} \right]
            \prod_{1 \leq j < k \leq n} 
            \left[ \frac{\Gamma \left( t z_j^{\pm 1} z_k^{\pm 1}; p,q \right)}{\Gamma \left( z_j^{\pm 1} z_k^{\pm 1}; p,q \right)} \right]
            \prod_{j=1}^n 
            \left[ \frac{\prod_{m=1}^6 \Gamma \left( t_m z_j^{\pm 1}; p,q \right)}{\Gamma \left( z_j^{\pm 2}; p,q \right)} \right] \\
            =&
            \prod_{j=1}^n \left[ \frac{\Gamma \left( t^j; p,q \right)}{\Gamma \left( t; p,q \right)} \right]
            \prod_{j=1}^n \left[ \prod_{1 \leq m < s \leq 6} \Gamma \left( t^{j-1} t_m t_s; p,q \right) \right],
            \numberthis
            \label{indexSpiridonovrewritten}
            \end{align*}
where $\Gamma$ is the standard elliptic gamma function. This $4d$ $USp(2N)$ duality corresponds in the statistical mechanics side to the following star-triangle type relation stated in \cite{Spiridonov:2010em}
            \begin{align*}            
            \int\limits_{[0, 2\pi]^n}
            [d \mathbf{u}] S(\mathbf{u}; t, p, q)
            W_{\eta - \alpha}(x, \mathbf{u}) W_{\alpha + \gamma}(y, \mathbf{u}) W_{\eta - \gamma}(w, \mathbf{u}) \hspace{4cm} \\
            =
            R(\alpha, \gamma, \eta; t, p, q)
            W_{\alpha}^t(y, w) W_{\eta - \alpha - \gamma}^t(x, w) W_{\gamma}^t(x, y),
            \numberthis
            \label{startriangletyperelationrewritten}
            \end{align*}
            where $S({\bf u}; t, p, q)$ and $R(\alpha, \gamma, \eta; t, p, q)$ are the interaction and normalization factors, respectively, while 
$W_{\eta - \alpha}(x, {\bf u})$ and $W_{\alpha}^t(y, w)$ are the two types of associated Boltzmann weights. The Boltzmann weights for this model are explicitly calculated\footnote{Calculations here contain a slightly different definition of the measure and of the interaction and normalization factors from those in reference \cite{Spiridonov:2010em}.}
by comparing the supersymmetric duality 
(\ref{indexSpiridonovrewritten}) and the STR type expression (\ref{startriangletyperelationrewritten}). To this end, consider the following definitions
            \begin{align*}
            t_1 &= \sqrt{pq} \ e^{\displaystyle \eta - \alpha + ix},  &  t_3 &= \sqrt{pq} \ e^{\displaystyle \alpha + \gamma + iy}, 
            &  t_5 &= \sqrt{pq} \ e^{\displaystyle \eta - \gamma + iw}, \\
            t_2 &= \sqrt{pq} \ e^{\displaystyle \eta - \alpha - ix},  &  t_4 &= \sqrt{pq} \ e^{\displaystyle \alpha + \gamma - iy}, 
            &  t_6 &= \sqrt{pq} \ e^{\displaystyle \eta - \gamma - iw}, \\
            z_j &= e^{iu_j},  &  pq &= t^{2n-2} \prod_{m=1}^6 t_m,  &  pq &= t^{-n+1} e^{-2\eta},
            \numberthis
            \label{expresionesfugacidadSpiridonov}
            \end{align*}
where equations involving $pq$ are the balancing condition and the definition of the crossing parameter $\eta$, in that order. Let's work explicitly both sides of equation (\ref{indexSpiridonovrewritten}). First, rewrite this equation by using (\ref{expresionesfugacidadSpiridonov}) as
            \begin{align*}
            &
            \int\limits_{[0, 2\pi]^n}
            \left[
            \frac{(p,p)^n_{\infty} (q,q)^n_{\infty}}{(4\pi)^n n!}
            \prod_{j=1}^n \left[ \frac{\Gamma \left( t; p,q \right)}{\Gamma \left( t^j; p,q \right)} \right]
            \prod_{j=1}^n du_j
            \right]
            \prod_{1 \leq j < k \leq n} 
            \left[ \frac{\Gamma \left( t e^{\pm iu_j} e^{\pm iu_k}; p,q \right)}{\Gamma \left( e^{\pm iu_j} e^{\pm iu_k}; p,q \right)} \right] \\
            & \times
            \prod_{j=1}^n \frac{1}{\Gamma \left( e^{\pm 2iu_j}; p,q \right)}
            \prod_{j=1}^n \prod_{m=1}^6 \Gamma \left( t_m e^{\pm iu_j}; p,q \right) \\
            =&
            \prod_{j=1}^n \prod_{1 \leq m < s \leq 6} \Gamma \left( t^{j-1} t_m t_s; p,q \right).
            \numberthis
            \label{indexSpiridonovrewrittentemporal}
            \end{align*}
            
By defining the measure $[d\mathbf{u}]$ and the interaction term $S(\mathbf{u}; t, p, q)$ as
            \begin{align*}
            [d\mathbf{u}] &= 
            \frac{(p,p)^n_{\infty} (q,q)^n_{\infty}}{(4\pi)^n n!}
            \prod_{j=1}^n \left[ \frac{\Gamma \left( t; p,q \right)}{\Gamma \left( t^j; p,q \right)} \right]
            \prod_{j=1}^n du_j, \\
            S(\mathbf{u}; t, p, q) &=
            \prod_{1 \leq j < k \leq n} 
            \left[ \frac{\Gamma \left( t e^{\pm iu_j} e^{\pm iu_k}; p,q \right)}{\Gamma \left( e^{\pm iu_j} e^{\pm iu_k}; p,q \right)} \right]
            \prod_{j=1}^n \frac{1}{\Gamma \left( e^{\pm 2iu_j}; p,q \right)},
            \numberthis
            \label{interactiontermSpiridonov}
            \end{align*}
where $\mathbf{u} = (u_1, \dots, u_n)$, it is possible to express (\ref{indexSpiridonovrewrittentemporal}) as
            \begin{align*}
            \int\limits_{[0, 2\pi]^n} [d \mathbf{u}] S(\mathbf{u}; t, p, q)
            \prod_{j=1}^n \prod_{m=1}^6 \Gamma \left( t_m e^{\pm iu_j}; p,q \right)
            =
            \prod_{j=1}^n \prod_{1 \leq m < s \leq 6} \Gamma \left( t^{j-1} t_m t_s; p,q \right).
            \numberthis
            \label{indexSpiridonovrewrittenchido}
            \end{align*}
            
For the left hand of (\ref{indexSpiridonovrewrittenchido}) we note that
            \begin{alignat*}{6}
            \prod_{m=1}^6 \Gamma \left( t_m e^{\pm iu_j}; p,q \right) 
            &=& 
            & \Gamma \left( \sqrt{pq} \ e^{\eta - \alpha + ix} e^{\pm iu_j}; p, q \right) 
              \Gamma \left( \sqrt{pq} \ e^{\eta - \alpha - ix} e^{\pm iu_j}; p, q \right) \\
            && \times &
              \Gamma \left( \sqrt{pq} \ e^{\alpha + \gamma + iy} e^{\pm iu_j}; p, q \right)
              \Gamma \left( \sqrt{pq} \ e^{\alpha + \gamma - iy} e^{\pm iu_j}; p, q \right) \\
            && \times &
              \Gamma \left( \sqrt{pq} \ e^{\eta - \gamma + iw} e^{\pm iu_j}; p, q \right)
              \Gamma \left( \sqrt{pq} \ e^{\eta - \gamma - iw} e^{\pm iu_j}; p, q \right) \\ \\
            &=& 
            & \Gamma \left( \sqrt{pq} \ e^{\eta - \alpha} e^{\pm ix} e^{\pm iu_j}; p, q \right)
              \Gamma \left( \sqrt{pq} \ e^{\alpha + \gamma} e^{\pm iy} e^{\pm iu_j}; p, q \right)\\
            && \times &
              \Gamma \left( \sqrt{pq} \ e^{\eta - \gamma} e^{\pm iw} e^{\pm iu_j}; p, q \right).
            \numberthis
            \label{ladoizquierdoSpiridonov}
            \end{alignat*}

Now, for the right hand side of (\ref{indexSpiridonovrewrittenchido}) one has
            \begin{alignat*}{6}
            \prod_{1 \leq m < s \leq 6} \Gamma \left( t^{j-1} t_m t_s; p,q \right)
            &=&
            & \bigg[ \Gamma \left( t^{j-1} (pq) \ e^{2(\eta - \alpha)}; p,q \right)
              \Gamma \left( t^{j-1} (pq) \ e^{2(\alpha + \gamma)}; p,q \right)  \\
            && \times &
              \Gamma \left( t^{j-1} (pq) \ e^{2(\eta - \gamma)}; p,q \right) \bigg]  \\
            && \times &
              \Gamma \left( t^{j-1} (pq) \ e^{\eta + \gamma} e^{\pm ix} e^{\pm iy}; p,q \right)  \\
            && \times &
              \Gamma \left( t^{j-1} (pq) \ e^{2\eta - \alpha - \gamma} e^{\pm ix} e^{\pm iw}; p,q \right)  \\
            && \times &
              \Gamma \left( t^{j-1} (pq) \ e^{\eta + \alpha} e^{\pm iy} e^{\pm iw}; p,q \right) \\ \\
            &=&
            & R(\alpha, \gamma, \eta; t, p, q) \\
            && \times & \Gamma \left( \sqrt{pq} \left( t^{j-\frac{n+1}{2}} \right) e^{\alpha} e^{\pm iy} e^{\pm iw}; p,q \right)  \\
            && \times & \Gamma \left( \sqrt{pq} \left( t^{j-\frac{n+1}{2}} \right) e^{\gamma} e^{\pm ix} e^{\pm iy}; p,q \right)  \\
            && \times & \Gamma \left( \sqrt{pq} \left( t^{j-\frac{n+1}{2}} \right) e^{\eta - \alpha - \gamma} e^{\pm ix} e^{\pm iw}; p,q \right),
            \numberthis
            \label{ladoderechoSpiridonov}
            \end{alignat*}
where the last equality introduced the normalization factor $R(\alpha, \gamma, \eta; t, p, q)$ as
            \begin{align}
            R(\alpha, \gamma, \eta; t, p, q) = \Gamma \left( t^{j-1} (pq) \ e^{2(\eta - \alpha)}; p,q \right)
                                               \Gamma \left( t^{j-1} (pq) \ e^{2(\alpha + \gamma)}; p,q \right)
                                               \Gamma \left( t^{j-1} (pq) \ e^{2(\eta - \gamma)}; p,q \right)
            \label{normalizationfactorSpiridonov}
            \end{align}
and it was used the equality
            \begin{align*}
            (pq) \left( t^{j-1} \right) e^{\eta + \alpha} = \sqrt{pq} \left( t^{j- \frac{n+1}{2}} \right) e^{\alpha},
            \end{align*}
which can be obtained from the last expression in (\ref{expresionesfugacidadSpiridonov}). 
Thus, by keeping (\ref{ladoizquierdoSpiridonov}) and (\ref{ladoderechoSpiridonov}) in mind, and taking the  definition of the Boltzmann weights as follows
            \begin{align*}
            W_{\eta-\alpha}(x,\mathbf{u}) &= \prod_{j=1}^n \Gamma \left( \sqrt{pq} \ e^{\eta - \alpha} e^{\pm ix} e^{\pm iu_j}; p, q \right), \\
            W_{\alpha+\gamma}(y,\mathbf{u}) &= \prod_{j=1}^n \Gamma \left( \sqrt{pq} \ e^{\alpha +\gamma} e^{\pm iy} e^{\pm iu_j}; p, q \right), \\
            W_{\eta-\gamma}(w,\mathbf{u}) &= \prod_{j=1}^n \Gamma \left( \sqrt{pq} \ e^{\eta - \gamma} e^{\pm iw} e^{\pm iu_j}; p, q \right), \\
            W_{\alpha}^t(y, w) &= \prod_{j=1}^n \Gamma \left( \sqrt{pq} \left( t^{j-\frac{n+1}{2}} \right) e^{\alpha} e^{\pm iy} e^{\pm iw}; p,q \right), \\
            W_{\eta-\alpha-\gamma}^t(x, w) &= \prod_{j=1}^n \Gamma \left( \sqrt{pq} \left( t^{j-\frac{n+1}{2}} \right) e^{\eta - \alpha - \gamma} e^{\pm ix} e^{\pm iw}; p,q \right), \\
            W_{\gamma}^t(x, y) &= \prod_{j=1}^n \Gamma \left( \sqrt{pq} \left( t^{j-\frac{n+1}{2}} \right) e^{\gamma} e^{\pm ix} e^{\pm iy}; p,q \right),
            \numberthis
            \label{BoltzmannweightsSpiridonov}
            \end{align*}
it is possible to rewrite (\ref{indexSpiridonovrewrittenchido}) exactly as (\ref{startriangletyperelationrewritten}), as desired. Note that the right hand side Boltzmann weights contain an extra parameter $t$ that is not present in the left hand side ones. This feature will be shared with our result in section \ref{CsakiSkibaSchmaltzduality}.

\section{STR type expressions for \texorpdfstring{$2d$ $\mathcal{N}=(0,2)$ $USp(2N)$}{} dualities}

Gauge theories $2d$ $\mathcal{N}=(0,2)$ are nicely described in \cite{Witten:1993yc,Melnikov:2019tpl}. The spectrum of these theories consist of three different multiplets, namely, the chiral multiplet $\Phi$ with one chiral fermion $\Psi_+$ and one complex scalar $\phi$, the vector multiplet $V$ with one fermion $\chi_-$ and gauge bosons $\left\{ v_{\alpha} \right\}_{\alpha=0,1}$, and the Fermi multiplet $\Lambda$ with one chiral spinor $\lambda_-$, a holomorphic function $E$ of the chiral superfields $\Phi_i$ and an auxiliary field $G$.
The contributions to the index (elliptic flavored genus in the NS-NS sector) of $2d$ $\mathcal{N}=(0,2)$ theories coming from chiral, vector and Fermi multiplets are calculated in \cite{Gadde:2013wq} and they are given in terms of the Jacobi theta functions.

\subsection{\texorpdfstring{$2d$ $\mathcal{N}=(0,2)$ $USp(2N)$}{} Intriligator-Pouliot duality}

In this subsection we analyse $2d$ $\mathcal{N}=(0,2)$ $USp(2N)$ Intriligator-Pouliot duality and we build star-triangle type relations for different values of $N$. As stated in \cite{Gadde:2015wta}, this duality comes from dimensional reduction on $\mathbb{S}^2$ of $4d$ $\mathcal{N}=1$ $USp(2N)$ confining Intriligator-Pouliot duality (this one, first studied in \cite{Csaki:1996eu}, is the $USp(2N)$ version of Seiberg duality). The $2d$ $\mathcal{N}=(0,2)$ $USp(2N)$ Intriligator-Pouliot duality is realized between a $USp(2N)$ gauge theory with $2N+2$ chiral multiplets in the fundamental representation, and a Laudau-Ginzburg model with $(N+1)(2N+1)$ chiral multiplets and a Fermi multiplet. The elliptic flavoured genera expression for the duality of these $2d$ $\mathcal{N}=(0,2)$ $USp(2N)$ supersymmetric quiver gauge theories is, from \cite{Sacchi:2020pet}\footnote{As stated in this reference, equality (\ref{indexIntriligatorPouliot}) can be tested perturbatively in variable $q$.},
            \begin{align}
            \int \frac{d \bar{z}_N}{\prod_{i=1}^N \prod_{a=1}^{2N+2} \theta \left( s u_a z_i^{\pm 1}; q \right)}
            =
            \frac{\theta \left( q s^{-2(N+1)}; q \right)}{\prod_{1 \leq a < b \leq 2N+2} \theta \left( s^2 u_a u_b; q \right)},
            \label{indexIntriligatorPouliot}
            \end{align}
where
            \begin{align}
            d \bar{z}_N = \frac{(q;q)_{\infty}^{2N}}{N! (4 \pi)^N} 
            \prod_{i=1}^N \left[ \frac{dz_i}{i z_i} \theta \left( z_i^{\pm 2}; q \right) \right]
            \prod_{1 \leq i < j \leq N} \theta \left( z_i^{\pm 1} z_j^{\pm 1}; q \right)
            \label{medidaIntriligatorPouliot}
            \end{align}
is the measure associated with $USp(2N)$. Here, $\{u_a\}_{a=1, \dots, 2N+2}$ and $\{s\}$ are the sets of fugacities associated with the global symmetry group $SU(2N+2)_u \times U(1)_s$ of these theories. Thus, to match with star-triangle type relations the three spectral parameters $\alpha$, $\beta$ and $\gamma$ have to be distributed into $N+1$ pairs of fugacities as in the subsequent expressions (\ref{expresionesfugacidadIntriligatorPouliotN=1}), (\ref{expresionesfugacidadIntriligatorPouliot}), (\ref{expresionesfugacidadIntriligatorPouliotN=3}), (\ref{expresionesfugacidadIntriligatorPouliotN=4}), (\ref{expresionesfugacidadIntriligatorPouliotN=5}) and (\ref{expresionesfugacidadIntriligatorPouliotN=3k+2}). There would be then $4N(N+1)$ theta functions having spin variables $x_i$, $i=1, \dots, N+1$, in the left hand side of (\ref{indexIntriligatorPouliot}) while there will be $2N(N+1)$ in the right hand one. It will also be useful to define the following expressions
            \begin{align*}
            z_i &= e^{i\Omega_i}, \numberthis \label{definicionzNgeneral} \\
            \left[ d\mathbf{\Omega} \right] &= \prod_{i=1}^N d\Omega_i, \numberthis \label{medidaNgeneral} \\
            S \left( \mathbf{\Omega}; q \right) &= \prod_{i=1}^N \theta \left( e^{\pm 2i\Omega_i}; q \right)
            \prod_{1 \leq i < j \leq N} \theta \left( e^{\pm i\Omega_i} e^{\pm i\Omega_j}; q \right), 
            \numberthis \label{interactionfactorNgeneral}
            \end{align*}
where (\ref{interactionfactorNgeneral}) will stand for the interaction factor for any value of $N$, so that (\ref{indexIntriligatorPouliot}) can be written as
            \begin{align}
            \int \frac{ \left[d\mathbf{\Omega}\right] S \left( \mathbf{\Omega}; q \right) }
                      { \prod_{i=1}^N \prod_{a=1}^{2N+2} \theta \left( s u_a e^{\pm i\Omega_i}; q \right) }
            =
            \frac{N! (4 \pi)^N}{(q;q)_{\infty}^{2N}}
            \left[ 
            \frac{\theta \left( q s^{-2(N+1)}; q \right)}{\prod_{1 \leq a < b \leq 2N+2} \theta \left( s^2 u_a u_b; q \right)}
            \right].
            \label{indexIntriligatorPouliotnuevoscasosrewritten}
            \end{align}

Define also the crossing parameter $\eta$ as
            \begin{align}
            \eta = \alpha + \beta + \gamma
            \label{crossingparameterNgeneral}
            \end{align}
and the following notation that will be used throughout the whole work
            \begin{align}
            \theta \left( a e^{\pm b}; q \right) = \theta \left( a e^{b}; q \right) \theta \left( a e^{-b}; q \right).
            \label{propiedadthetaIntriligatorPouliotNgeneral}
            \end{align}
            
In the following subsections we obtain STR type expressions for $2d$ $\mathcal{N}=(0,2)$ $USp(2N)$ Intriligator-Pouliot duality (\ref{indexIntriligatorPouliot}).
In section \ref{Case \texorpdfstring{$N=1$}{}} we obtain an expression analogous to the so called triangle identity identified in \cite{Kels:2018xge} in the context of Yang–Baxter/$3D$-consistency correspondence. 
In sections \ref{Case \texorpdfstring{$N=3$}{}} and  \ref{Case \texorpdfstring{$N=4$}{}} we observe a similarity with the asymmetric form of the star-triangle relation \cite{Kels:2019ktt,Kels:2018xge,Kels:2020zjn}. Finally, in sections \ref{Case \texorpdfstring{$N=2$}{}}, \ref{Case \texorpdfstring{$N=5$}{}} and \ref{Case \texorpdfstring{$N=3k+2$}{}} an attempt to build an STR type expression is carried out.

\subsubsection{Case \texorpdfstring{$N=1$}{}} \label{Case \texorpdfstring{$N=1$}{}}

The analysis of $N=1$ case is interesting because duality (\ref{indexIntriligatorPouliot}) reduces, by using (\ref{medidaIntriligatorPouliot}), to
            \begin{align*}
            \frac{(q;q)_{\infty}^{2}}{4 \pi}
            \int 
            \left[
            \frac{dz}{i z} \theta \left( z^{\pm 2}; q \right)
            \right]
            \prod_{a=1}^{4} \big[ \theta \left( s u_a z^{\pm 1}; q \right) \big]^{-1}
            =
            \theta \left( q s^{-4}; q \right) 
            \prod_{1 \leq a < b \leq 4} \big[ \theta \left( s^2 u_a u_b; q \right) \big]^{-1},
            \numberthis
            \label{indexIntriligatorPouliotN=1}
            \end{align*}
which is precisely $2d$ $\mathcal{N}=(0,2)$ $SU(2)$ duality considered in \cite{Dedushenko:2017osi,Gadde:2015wta} between a $SU(2)$ gauge theory with $4$ chiral 
multiplets in the fundamental representation and a Landau-Ginzburg model with $6$ chiral multiplets and a Fermi multiplet.
            
By using (\ref{propiedadthetaIntriligatorPouliotNgeneral}) and defining the following relations between fugacities, spectral parameters and spin variables as
            \begin{align*}
            u_1 &= s^{-1}e^{-\alpha + ix_1},  &  u_3 &= s^{-1}e^{-\beta + ix_2}, \\
            u_2 &= s^{-1}e^{-\alpha - ix_1},  &  u_4 &= s^{-1}e^{-\beta - ix_2},
            \numberthis
            \label{expresionesfugacidadIntriligatorPouliotN=1}
            \end{align*}
and the balancing condition as
            \begin{align}
            \prod_{a=1}^4 u_a = \frac{1}{q},
            \label{balancingconditionIntriligatorPouliotN=1}
            \end{align}
it is possible to write some factors in (\ref{indexIntriligatorPouliotN=1}) as
            \begin{align*}
            \prod_{a=1}^{4} \theta \left( s u_a z^{\pm 1}; q \right) 
            &=
            \theta \left( e^{-\alpha + ix_1} e^{\pm i\Omega}; q \right) \theta \left( e^{-\alpha - ix_1} e^{\pm i\Omega}; q \right) 
            \theta \left( e^{-\beta + ix_2} e^{\pm i\Omega}; q \right) \theta \left( e^{-\beta - ix_2} e^{\pm i\Omega}; q \right) \\
            &=
            \theta \left( e^{-\alpha} e^{\pm ix_1} e^{\pm i\Omega}; q \right) \theta \left( e^{-\beta} e^{\pm ix_2} e^{\pm i\Omega}; q \right)
            \label{ladoizquierdoIntriligatorPouliotN=1}
            \numberthis
            \end{align*}
and
            \begin{alignat*}{6}
            \prod_{1 \leq a < b \leq 4} \theta \left( s^2 u_a u_b; q \right)
            &=& &
            \theta \left( e^{-2\alpha}; q \right) \theta \left( e^{-2\beta}; q \right) 
            \theta \left( e^{-(\alpha + \beta) + ix_1 + ix_2}; q \right)  \\
            & &\times&
            \theta \left( e^{-(\alpha + \beta) + ix_1 - ix_2}; q \right)
            \theta \left( e^{-(\alpha + \beta) - ix_1 + ix_2}; q \right) \theta \left( e^{-(\alpha + \beta) - ix_1 - ix_2}; q \right) \\ \\
            &=& &
            \theta \left( e^{-2\alpha}; q \right) \theta \left( e^{-2\beta}; q \right) 
            \theta \left( e^{-(\alpha + \beta)} e^{\pm ix_1} e^{\pm ix_2}; q \right).
            \label{ladoderechoIntriligatorPouliotN=1}
            \numberthis
            \end{alignat*}
            
Also, note that the balancing condition (\ref{balancingconditionIntriligatorPouliotN=1}) implies the following relation
            \begin{align}
            qs^{-4} = e^{2(\alpha + \beta)}.
            \label{saledebalancingconditionIntriligatorPouliotN=1}
            \end{align}

Now, by using Eqs. (\ref{definicionzNgeneral}), (\ref{ladoizquierdoIntriligatorPouliotN=1}), (\ref{ladoderechoIntriligatorPouliotN=1}) and (\ref{saledebalancingconditionIntriligatorPouliotN=1}), the index duality (\ref{indexIntriligatorPouliotN=1}) can be rewritten as
            \begin{align*}
            \frac{(q;q)_{\infty}^{2}}{4 \pi}
            \int 
            \left[ d\mathbf{\Omega} \right]
            \bigg[ \theta \left( e^{\pm 2i\Omega}; q \right) \bigg]
            \bigg[
            \theta \left( e^{-\alpha} e^{\pm ix_1} e^{\pm i\Omega}; q \right) \theta \left( e^{-\beta} e^{\pm ix_2} e^{\pm i\Omega}; q \right) 
            \bigg]^{-1} \\
            =
            \theta \left( e^{2(\alpha + \beta)}; q \right) 
            \bigg[ 
            \theta \left( e^{-2\alpha}; q \right) \theta \left( e^{-2\beta}; q \right) 
            \theta \left( e^{-(\alpha + \beta)} e^{\pm ix_1} e^{\pm ix_2}; q \right) 
            \bigg]^{-1}.
            \numberthis
            \label{indexIntriligatorPouliotN=1rewritten}
            \end{align*}

The definition of the interaction and normalization factors, $S\left(\mathbf{\Omega}; q\right)$ and $R(\alpha, \beta)$, respectively, as
            \begin{align*}
            S\left(\mathbf{\Omega}; q\right) &=
            \left( e^{\pm 2i\Omega}; q \right), \\
            R(\alpha, \beta) &= \frac{1! (4 \pi)^1}{(q;q)_{\infty}^{2(1)}}
            \theta \left( e^{2(\alpha + \beta)}; q \right)
            \Big[ \theta \left( e^{-2\alpha}; q \right) \theta \left( e^{-2\beta}; q \right) \Big]^{-1},
            \numberthis
            \label{normalizationfactorIntriligatorPouliotN=1}
            \end{align*}
and the Boltzmann weights as
            \begin{align*}
            W_{\alpha} \left( x_1, \mathbf{\Omega} \right)    &= \big[ \theta \left( e^{-\alpha} e^{\pm ix_1} e^{\pm i\Omega}; q \right) \big]^{-1}, \\
            W_{\beta} \left( x_2, \mathbf{\Omega} \right)     &= \big[ \theta \left( e^{-\beta} e^{\pm ix_2} e^{\pm i\Omega}; q \right) \big]^{-1}, \\
            W_{\alpha + \beta} \left( x_1, x_2 \right) &= \big[ \theta \left( e^{-(\alpha + \beta)} e^{\pm ix_1} e^{\pm ix_2}; q \right) \big]^{-1},
            \numberthis
            \label{BoltzmannweightsIntriligatorPouliotN=1}
            \end{align*}
allows us to write the index duality (\ref{indexIntriligatorPouliotN=1rewritten}) as
            \begin{align}
            \int \left[ d\mathbf{\Omega} \right] S\left(\mathbf{\Omega}; q\right)
            W_{\alpha} \left( x_1, \mathbf{\Omega} \right) W_{\beta} \left( x_2, \mathbf{\Omega} \right)
            =
            R(\alpha, \beta) W_{\alpha + \beta} \left( x_1, x_2 \right).
            \label{startriangletyperelationrewrittenIntriligatorN=1}
            \end{align}
            
Expression (\ref{startriangletyperelationrewrittenIntriligatorN=1}) is very interesting because its form is analogous to that of the triangle identity considered in the context of Yang–Baxter/$3D$-consistency correspondence \cite{Kels:2018xge}.

\subsubsection{Case \texorpdfstring{$N=2$}{}} \label{Case \texorpdfstring{$N=2$}{}}

Now let's consider the case $N=2$ for which (\ref{indexIntriligatorPouliot}) can be written, by using (\ref{medidaIntriligatorPouliot}), as
            \begin{align*}
            \int 
            \left[
            \frac{(q;q)_{\infty}^{2(2)}}{2! (4 \pi)^2} 
            \left[ \prod_{i=1}^2 \frac{dz_i}{i z_i} \right] \prod_{i=1}^2\theta \left( z_i^{\pm 2}; q \right)
            \prod_{1 \leq i < j \leq 2} \theta \left( z_i^{\pm 1} z_j^{\pm 1}; q \right)
            \right]
            \prod_{i=1}^2 \prod_{a=1}^{6} \bigg[ \theta \left( s u_a z_i^{\pm 1}; q \right) \bigg]^{-1} \\
            =
            \theta \left( q s^{-6}; q \right) \prod_{1 \leq a < b \leq 6} \bigg[ \theta \left( s^2 u_a u_b; q \right) \bigg]^{-1}.
            \numberthis
            \label{indexIntriligatorPouliotN=2}
            \end{align*}
           
In order to work explicitly both sides of expression (\ref{indexIntriligatorPouliotN=2}) define the following relations
            \begin{align*}
            u_1 &= s^{-1}e^{-\alpha + ix_1},  &  u_3 &= s^{-1}e^{-\beta + ix_2},  &  u_5 &= s^{-1}e^{-\gamma + ix_3}, \\
            u_2 &= s^{-1}e^{-\alpha - ix_1},  &  u_4 &= s^{-1}e^{-\beta - ix_2},  &  u_6 &= s^{-1}e^{-\gamma - ix_3},
            \numberthis
            \label{expresionesfugacidadIntriligatorPouliot}
            \end{align*}
as well as the balancing condition
            \begin{align}
            \prod_{a=1}^6 u_a = \frac{1}{q}.
            \label{balancingconditionIntriligatorPouliot}
            \end{align}
            
By using (\ref{definicionzNgeneral}) and (\ref{expresionesfugacidadIntriligatorPouliot}), the left and right hand sides of (\ref{indexIntriligatorPouliotN=2}) can be rewritten as  
            \begin{alignat*}{6}
            \prod_{i=1}^2 \prod_{a=1}^6 \theta \left( s u_a z_i^{\pm 1}; q \right) 
            =
            \prod_{i=1}^2 \bigg[ &
            \theta \left( s u_1 e^{\pm i\Omega_i}; q \right) 
            \theta \left( s u_2 e^{\pm i\Omega_i}; q \right)
            \theta \left( s u_3 e^{\pm i\Omega_i}; q \right) \\
            \times &
            \theta \left( s u_4 e^{\pm i\Omega_i}; q \right)
            \theta \left( s u_5 e^{\pm i\Omega_i}; q \right) 
            \theta \left( s u_6 e^{\pm i\Omega_i}; q \right) \bigg] \\
            =
            \prod_{i=1}^2 \bigg[ &
            \theta\left(e^{-\alpha} e^{ix_1} e^{\pm i\Omega_i}; q\right) 
            \theta\left(e^{-\alpha} e^{-ix_1} e^{\pm i\Omega_i}; q\right) \\
            \times &
            \theta\left(e^{-\beta} e^{ix_2} e^{\pm i\Omega_i}; q\right) 
            \theta\left(e^{-\beta} e^{-ix_2} e^{\pm i\Omega_i}; q\right) \\
            \times &
            \theta\left(e^{-\gamma} e^{ix_3} e^{\pm i\Omega_i}; q\right) 
            \theta\left(e^{-\gamma} e^{-ix_3} e^{\pm i\Omega_i}; q\right) \bigg] \\
            =
            \prod_{i=1}^2 \bigg[ &
            \theta\left(e^{-\alpha} e^{\pm ix_1} e^{\pm i\Omega_i}; q\right)
            \theta\left(e^{-\beta} e^{\pm ix_2} e^{\pm i\Omega_i}; q\right)
            \theta\left(e^{-\gamma} e^{\pm ix_3} e^{\pm i\Omega_i}; q\right) \bigg],
            \numberthis
            \label{ladoizquierdoIntriligatorPouliot}
            \end{alignat*}
and
            \begin{alignat*}{6}
            \prod_{1 \leq a < b \leq 6} \theta \left( s^2 u_a u_b; q \right) &=& &
            \theta\left( s^2 u_1 u_2; q \right) \theta\left( s^2 u_1 u_3; q \right) \theta\left( s^2 u_1 u_4; q \right)
            \theta\left( s^2 u_1 u_5; q \right) \theta\left( s^2 u_1 u_6; q \right) \\
             & &\times&
            \theta\left( s^2 u_2 u_3; q \right) \theta\left( s^2 u_2 u_4; q \right) \theta\left( s^2 u_2 u_5; q \right)
            \theta\left( s^2 u_2 u_6; q \right) \theta\left( s^2 u_3 u_4; q \right) \\
             & &\times&
            \theta\left( s^2 u_3 u_5; q \right) \theta\left( s^2 u_3 u_6; q \right) \theta\left( s^2 u_4 u_5; q \right)
            \theta\left( s^2 u_4 u_6; q \right) \theta\left( s^2 u_5 u_6; q \right) \\ \\
             &=& &
            \bigg[ \theta\left( e^{-2\alpha}; q \right) \theta\left( e^{-2\beta}; q \right) \theta\left( e^{-2\gamma}; q \right) \bigg] \\
             & &\times&
            \bigg[ \theta\left( e^{-(\alpha + \beta) + ix_1 + ix_2}; q \right) \theta\left( e^{-(\alpha + \beta) + ix_1 - ix_2}; q \right) 
                   \theta\left( e^{-(\alpha + \beta) - ix_1 + ix_2}; q \right)  \\
             & &\times&
            \theta\left( e^{-(\alpha + \beta) - ix_1 - ix_2}; q \right) \bigg] \\
             & &\times&
            \bigg[ \theta\left( e^{-(\alpha + \gamma) + ix_1 + ix_3}; q \right) \theta\left( e^{-(\alpha + \gamma) + ix_1 - ix_3}; q \right)
                   \theta\left( e^{-(\alpha + \gamma) - ix_1 + ix_3}; q \right)  \\
             & &\times&
            \theta\left( e^{-(\alpha + \gamma) - ix_1 - ix_3}; q \right) \bigg] \\
             & &\times&
            \bigg[ \theta\left( e^{-(\beta + \gamma) + ix_2 + ix_3}; q \right) \theta\left( e^{-(\beta + \gamma) + ix_2 - ix_3}; q \right)
                   \theta\left( e^{-(\beta + \gamma) - ix_2 + ix_3}; q \right) \\
             & &\times&
            \theta\left( e^{-(\beta + \gamma) - ix_2 - ix_3}; q \right) \bigg] \\
             &=& &
            \bigg[ \theta\left( e^{-2\alpha}; q \right) \theta\left( e^{-2\beta}; q \right) \theta\left( e^{-2\gamma}; q \right) \bigg] \\
             & &\times& \bigg[
            \theta\left( e^{-(\alpha + \beta)} e^{\pm ix_1} e^{\pm ix_2}; q \right) 
            \theta\left( e^{-(\alpha + \gamma)} e^{\pm ix_1} e^{\pm ix_3}; q \right)
            \theta\left( e^{-(\beta + \gamma)} e^{\pm ix_2} e^{\pm ix_3}; q \right) \bigg],
            \numberthis
            \label{ladoderechoIntriligatorPouliot}
            \end{alignat*}
respectively. It is also important to remark that balancing condition (\ref{balancingconditionIntriligatorPouliot}) implies the relation
            \begin{align}
            q s^{-6} = e^{2(\alpha + \beta + \gamma)}.
            \label{saledebalancingconditionIntriligatorPouliot}
            \end{align}

Then, by using Eqs. (\ref{ladoizquierdoIntriligatorPouliot}), (\ref{ladoderechoIntriligatorPouliot}) and (\ref{saledebalancingconditionIntriligatorPouliot}), the index duality (\ref{indexIntriligatorPouliotN=2}) can be rewritten as
            \begin{alignat*}{6}
            & & & \int 
            \left[ \prod_{i=1}^2 d\Omega_i \right] \left[ \prod_{i=1}^2\theta \left( e^{\pm 2i \Omega_i}; q \right)
            \prod_{1 \leq i < j \leq 2} \theta \left( e^{\pm i\Omega_i} e^{\pm i\Omega_j}; q \right)
            \right]  \\
            & &\times& \prod_{i=1}^2 \bigg[
            \theta \left(e^{-\alpha} e^{\pm ix_1} e^{\pm i\Omega_i}; q\right)
            \theta \left(e^{-\beta} e^{\pm ix_2} e^{\pm i\Omega_i}; q\right)
            \theta \left(e^{-\gamma} e^{\pm ix_3} e^{\pm i\Omega_i}; q\right) \bigg]^{-1} \\
            &=& &
            \frac{2! (4 \pi)^2}{(q;q)_{\infty}^{4}}
            \bigg[ \theta \left( e^{-2(\alpha + \beta + \gamma)}; q \right) \bigg]
            \bigg[ \theta \left( e^{-2\alpha}; q \right) 
                   \theta \left( e^{-2\beta}; q \right) 
                   \theta \left( e^{-2\gamma}; q \right) \bigg]^{-1}  \\
            & &\times&
            \bigg[
            \theta \left( e^{-(\alpha + \beta)} e^{\pm ix_1} e^{\pm ix_2}; q \right) 
            \theta \left( e^{-(\alpha + \gamma)} e^{\pm ix_1} e^{\pm ix_3}; q \right)
            \theta \left( e^{-(\beta + \gamma)} e^{\pm ix_2} e^{\pm ix_3}; q \right) 
            \bigg]^{-1}.
            \numberthis
            \label{indexIntriligatorPouliotN=2temporal}
            \end{alignat*}

Thus, by taking expression (\ref{medidaNgeneral}) as well as the definition of interaction and normalization factors, $S\left(\mathbf{\Omega}; q\right)$ and $R(\alpha, \beta, \gamma)$, respectively, as
            \begin{align*}
            S\left(\mathbf{\Omega}; q\right) &=
            \prod_{i=1}^2\theta \left( e^{\pm 2i\Omega_i}; q \right)
            \prod_{1 \leq i < j \leq 2} \theta \left( e^{\pm i\Omega_i} e^{\pm i\Omega_j}; q \right), \\
            R(\alpha, \beta, \gamma) &= \frac{2! (4 \pi)^2}{(q;q)_{\infty}^{4}}
            \bigg[ \theta \left( e^{2(\alpha + \beta + \gamma)}; q \right) \bigg]
            \bigg[ \theta \left( e^{-2\alpha}; q \right) 
                   \theta \left( e^{-2\beta}; q \right) 
                   \theta \left( e^{-2\gamma}; q \right) \bigg]^{-1},
            \numberthis
            \label{normalizationfactorIntriligatorPouliotN=2}
            \end{align*}
the crossing parameter as (\ref{crossingparameterNgeneral}) and the Boltzmann weights as
            \begin{align*}
            W_{\alpha}^{\pm} \left( x_1, \mathbf{\Omega} \right) 
            &= \prod_{i=1}^2 \big[ \theta \left(e^{-\alpha} e^{\pm ix_1} e^{\pm i\Omega_i}; q\right) \big]^{-1}, \\
            W_{\beta}^{\pm} \left( x_2, \mathbf{\Omega} \right)
            &= \prod_{i=1}^2 \big[ \theta \left(e^{-\beta} e^{\pm ix_2} e^{\pm i\Omega_i}; q\right) \big]^{-1}, \\
            W _{\gamma}^{\pm} \left( x_3, \mathbf{\Omega} \right)
            &= \prod_{i=1}^2 \big[ \theta \left(e^{-\gamma} e^{\pm ix_3} e^{\pm i\Omega_i}; q\right) \big]^{-1}, \\ \\
            W_{\eta-\gamma}(x_1, x_2, \_) &= \big[ \theta \left(e^{-(\alpha + \beta)} e^{\pm ix_1} e^{\pm ix_2}; q\right) \big]^{-1}, \\
            W_{\eta-\beta}(x_1, \_, x_3) &= \big[ \theta \left(e^{-(\alpha + \gamma)} e^{\pm ix_1} e^{\pm ix_3}; q\right) \big]^{-1}, \\
            W_{\eta-\alpha}(\_, x_2, x_3) &= \big[ \theta \left(e^{-(\beta + \gamma)} e^{\pm ix_2} e^{\pm ix_3}; q\right) \big]^{-1},
            \numberthis
            \label{BoltzmannweightsIntriligatorPouliotN=2}
            \end{align*}
it is possible to rewrite (\ref{indexIntriligatorPouliotN=2temporal}) exactly as the following STR type expression
            \begin{align*}
            \int \left[d\mathbf{\Omega}\right] S\left(\mathbf{\Omega}; q\right)
            W_{\alpha}^{\pm} \left( x_1, \mathbf{\Omega} \right) 
            W_{\beta}^{\pm} \left( x_2, \mathbf{\Omega} \right)
            W _{\gamma}^{\pm} \left( x_3, \mathbf{\Omega} \right) \hspace{6cm} \\
            =
            R(\alpha, \beta, \gamma)
            W_{\eta-\alpha}(x_2, x_3) W_{\eta-\beta}(x_1, x_3) W_{\eta-\gamma}(x_1, x_2),
            \numberthis
            \label{startriangletyperelationrewrittenIntriligator}
            \end{align*}
which resembles an STR expression because of the distribution of the spin variables as well as the spectral parameters in both sides of the relation despite the different definition of left and right hand side Boltzmann weights.
            
\subsubsection{Case \texorpdfstring{$N=3$}{}} \label{Case \texorpdfstring{$N=3$}{}}

The analysis is done in a similar way to those of the previous subsections. For this case it is convenient to define the following expressions
            \begin{align*}
            u_1 &= s^{-1}e^{-\alpha + ix_1},  &  u_3 &= s^{-1}e^{-\alpha + ix_2},  &  u_5 &= s^{-1}e^{-\beta + ix_3},  & 
            u_7 &= s^{-1}e^{-\gamma + ix_4}, \\ 
            u_2 &= s^{-1}e^{-\alpha - ix_1},  &  u_4 &= s^{-1}e^{-\alpha - ix_2},  &  u_6 &= s^{-1}e^{-\beta - ix_3},  &  
            u_8 &= s^{-1}e^{-\gamma - ix_4},  
            \numberthis
            \label{expresionesfugacidadIntriligatorPouliotN=3}
            \end{align*}
and the balancing condition
            \begin{align}
            \prod_{a=1}^8 u_a = \frac{1}{q},
            \label{balancingconditionIntriligatorPouliotN=3}
            \end{align}
which in turn implies
            \begin{align*}
            q s^{-8} = e^{2( 2\alpha + \beta + \gamma )}.
            \label{saledebalancingconditionIntriligatorPouliotN=3}
            \numberthis
            \end{align*}
            
By using the same notation that we followed in the previous subsections it is possible to write
            \begin{align*}
            \prod_{i=1}^{N} \prod_{a=1}^{2N+2} \theta \left( s u_a z_i^{\pm 1}; q \right) 
            =
            \prod_{i=1}^3 \Big[ &
            \theta \left( e^{-\alpha + ix_1} e^{\pm i\Omega_i}; q \right) \theta \left( e^{-\alpha - ix_1} e^{\pm i\Omega_i}; q \right) 
            \theta \left( e^{-\alpha + ix_2} e^{\pm i\Omega_i}; q \right) \\
            \times &
            \theta \left( e^{-\alpha - ix_2} e^{\pm i\Omega_i}; q \right)
            \theta \left( e^{-\beta + ix_3} e^{\pm i\Omega_i}; q \right) \theta \left( e^{-\beta - ix_3} e^{\pm i\Omega_i}; q \right) \\
            \times &
            \theta \left( e^{-\gamma + ix_4} e^{\pm i\Omega_i}; q \right) \theta \left( e^{-\gamma - ix_4} e^{\pm i\Omega_i}; q \right) \Big] \\
            =
            \prod_{i=1}^3 \Big[ &
            \theta \left( e^{-\alpha} e^{\pm ix_1} e^{\pm i\Omega_i}; q \right) 
            \theta \left( e^{-\alpha} e^{\pm ix_2} e^{\pm i\Omega_i}; q \right) \\
            \times &
            \theta \left( e^{-\beta} e^{\pm ix_3} e^{\pm i\Omega_i}; q \right)
            \theta \left( e^{-\gamma} e^{\pm ix_4} e^{\pm i\Omega_i}; q \right) \Big]
            \label{ladoizquierdoIntriligatorPouliotN=3}
            \numberthis
            \end{align*}
and
            \begin{alignat*}{6}
            \prod_{1 \leq a < b \leq 2N+2} \theta \left( s^2 u_a u_b; q \right) \times
            &=& &
            \Big[ \theta \left( e^{-2\alpha}; q \right) \Big]^2 
            \theta \left( e^{-2\beta}; q \right) 
            \theta \left( e^{-2\gamma}; q \right) \\
            & &\times&
            \theta \left( e^{-2\alpha \pm ix_1 \pm ix_2}; q \right) \theta \left( e^{-(\alpha + \beta) \pm ix_1 \pm ix_3}; q \right)
            \theta \left( e^{-(\alpha + \gamma) \pm ix_1 \pm ix_4}; q \right) \\
            & &\times&
            \theta \left( e^{-(\alpha + \beta) \pm ix_2 \pm ix_3}; q \right) 
            \theta \left( e^{-(\alpha + \gamma) \pm ix_2 \pm ix_4}; q \right) \\
            & &\times&
            \theta \left( e^{-(\beta + \gamma) \pm ix_3 \pm ix_4}; q \right).
            \label{ladoderechoIntriligatorPouliotN=3}
            \numberthis
            \end{alignat*}

We use again (\ref{medidaNgeneral}), the interaction factor (\ref{interactionfactorNgeneral}), and the normalization factor $R(\alpha, \beta, \gamma)$ given by
            \begin{align}
            R(\alpha, \beta, \gamma)
            =
            \frac{3! (4 \pi)^3}{(q;q)_{\infty}^{2(3)}}
            \theta \left( e^{2( 2\alpha + \beta + \gamma )}; q \right) 
            \Bigg[
            \big[ \theta \left( e^{-2\alpha}; q \right) \big]^2 
            \theta \left( e^{-2\beta}; q \right) \theta \left( e^{-2\gamma}; q \right)
            \Bigg]^{-1}
            \label{normalizationfactorIntriligatorPouliotN=3}
            \end{align}
as well as the crossing parameter (\ref{crossingparameterNgeneral}) and expressions (\ref{ladoizquierdoIntriligatorPouliotN=3}) and (\ref{ladoderechoIntriligatorPouliotN=3}) to write (\ref{indexIntriligatorPouliotnuevoscasosrewritten}) as
            \begin{alignat*}{6}
            \int \left[d\mathbf{\Omega}\right] S \left( \mathbf{\Omega}; q \right)
            \prod_{i=1}^3 \Bigg[
            & \theta \left( e^{-\alpha} e^{\pm ix_1} e^{\pm i\Omega_i}; q \right) 
            \theta \left( e^{-\alpha} e^{\pm ix_2} e^{\pm i\Omega_i}; q \right) \\
            \times &
            \theta \left( e^{-\beta} e^{\pm ix_3} e^{\pm i\Omega_i}; q \right)
            \theta \left( e^{-\gamma} e^{\pm ix_4} e^{\pm i\Omega_i}; q \right)
            \Bigg]^{-1} \\
            =
            R(\alpha, \beta, \gamma)
            \Bigg[ &
            \theta \left( e^{-2\alpha \pm ix_1 \pm ix_2}; q \right) 
            \theta \left( e^{-(\eta - \gamma) \pm ix_1 \pm ix_3}; q \right)
            \theta \left( e^{-(\eta - \beta) \pm ix_1 \pm ix_4}; q \right) \\
            \times &
            \theta \left( e^{-(\eta - \gamma) \pm ix_2 \pm ix_3}; q \right) 
            \theta \left( e^{-(\eta - \beta) \pm ix_2 \pm ix_4}; q \right)
            \theta \left( e^{-(\eta - \alpha) \pm ix_3 \pm ix_4}; q \right)
            \Bigg]^{-1}.
            \numberthis
            \label{indexIntriligatorPouliotnuevoscasosrewrittenN=3}
            \end{alignat*}

The last equation is quite suggestive and definition of the Boltzmann weights as
            \begin{align*}
            W_{\alpha} \left( x_1, x_2, \mathbf{\Omega} \right)
            &= 
                                  \frac{ \displaystyle \prod_{i=1}^3
                                  \big[ \theta \left( e^{-\alpha} e^{\pm ix_1} e^{\pm i\Omega_i}; q \right) 
                                  \theta \left( e^{-\alpha} e^{\pm ix_2} e^{\pm i\Omega_i}; q \right) \big]^{-1}}
                                  { \big[ \theta \left( e^{-2\alpha \pm ix_1 \pm ix_2}; q \right) \big]^{-1}}, \\ \\
            V_{\beta} \left( x_3, \mathbf{\Omega} \right)
            &= \prod_{i=1}^3 \big[ \theta \left( e^{-\beta} e^{\pm ix_3} e^{\pm i\Omega_i}; q \right) \big]^{-1}, \\
            V_{\gamma} \left( x_4, \mathbf{\Omega} \right)
            &= \prod_{i=1}^3 \big[ \theta \left( e^{-\gamma} e^{\pm ix_4} e^{\pm i\Omega_i}; q \right) \big]^{-1}, \\ \\
            \overline{W}_{\eta - \alpha} \left( \_, \_, x_3, x_4 \right)
            &= \big[ \theta \left( e^{-(\eta - \alpha) \pm ix_3 \pm ix_4}; q \right) \big]^{-1}, \\ \\
            \overline{V}_{\eta - \beta} \left( x_1, x_2, \_, x_4 \right)
            &= \big[ \theta \left( e^{-(\eta - \beta) \pm ix_1 \pm ix_4}; q \right) 
                     \theta \left( e^{-(\eta - \beta) \pm ix_2 \pm ix_4}; q \right) \big]^{-1}, \\ \\
            \overline{V}_{\eta - \gamma} \left( x_1, x_2, x_3, \_ \right)
            &= \big[ \theta \left( e^{-(\eta - \gamma) \pm ix_1 \pm ix_3}; q \right) 
                     \theta \left( e^{-(\eta - \gamma) \pm ix_2 \pm ix_3}; q \right) \big]^{-1},
            \numberthis
            \label{BoltzmannweightsIntriligatorPouliotN=3}
            \end{align*}
leads us to write down the index duality (\ref{indexIntriligatorPouliotnuevoscasosrewrittenN=3}) as
            \begin{align*}
            \int \left[d\mathbf{\Omega}\right] S \left( \mathbf{\Omega}; q \right) &
            W_{\alpha} \left( x_1, x_2, \mathbf{\Omega} \right)
            V_{\beta} \left( x_3, \mathbf{\Omega} \right)
            V_{\gamma} \left( x_4, \mathbf{\Omega} \right) \\
            = &
            R(\alpha, \beta, \gamma)
            \overline{W}_{\eta - \alpha} \left( \_, \_, x_3, x_4 \right)
            \overline{V}_{\eta - \beta} \left( x_1, x_2, \_, x_4 \right)
            \overline{V}_{\eta - \gamma} \left( x_1, x_2, x_3, \_ \right).
            \numberthis
            \label{startriangletyperelationrewrittenIntriligatorN=3}
            \end{align*}

Note that each side of (\ref{startriangletyperelationrewrittenIntriligatorN=3}) have the same definition for two Boltzmann weights while the third one is different. This feature resembles the graphical representation of the asymmetric form of the star-triangle relation \cite{Kels:2018xge}; unfortunately here there are more spin variables and the position of $V$ and $\overline{V}$ is not the same as in \cite{Kels:2019ktt,Kels:2018xge,Kels:2020zjn} for the asymmetric star-triangle relation in the context of Yang–Baxter/$3D$-consistency correspondence.

\subsubsection{Case \texorpdfstring{$N=4$}{}} \label{Case \texorpdfstring{$N=4$}{}}

The analysis is quite similar to that of the previous case. Define the expressions
            \begin{align*}
            u_1 &= s^{-1}e^{-\alpha + ix_1},  &  u_3 &= s^{-1}e^{-\alpha + ix_2},  &  u_5 &= s^{-1}e^{-\beta + ix_3},  & 
            u_7 &= s^{-1}e^{-\beta + ix_4},  &  u_9 &= s^{-1}e^{-\gamma + ix_5}, \\ 
            u_2 &= s^{-1}e^{-\alpha - ix_1},  &  u_4 &= s^{-1}e^{-\alpha - ix_2},  &  u_6 &= s^{-1}e^{-\beta - ix_3},  &  
            u_8 &= s^{-1}e^{-\beta - ix_4},  &  u_{10} &= s^{-1}e^{-\gamma - ix_5},
            \numberthis
            \label{expresionesfugacidadIntriligatorPouliotN=4}
            \end{align*}
and the balancing condition
            \begin{align}
            \prod_{a=1}^{10} u_a = \frac{1}{q},
            \label{balancingconditionIntriligatorPouliotN=4}
            \end{align}
which now implies
            \begin{align*}
            q s^{-10} = e^{2( 2\alpha + 2\beta + \gamma )}.
            \label{saledebalancingconditionIntriligatorPouliotN=4}
            \numberthis
            \end{align*}
            
This time, factors in the left and right hand sides of (\ref{indexIntriligatorPouliotnuevoscasosrewritten}) can be worked out as
            \begin{align*}
            \prod_{i=1}^{N} \prod_{a=1}^{2N+2} \theta \left( s u_a z_i^{\pm 1}; q \right)
            =
            \prod_{i=1}^4 \Big[ &
            \theta \left( e^{-\alpha} e^{\pm ix_1} e^{\pm i\Omega_i}; q \right) 
            \theta \left( e^{-\alpha} e^{\pm ix_2} e^{\pm i\Omega_i}; q \right) \\
            \times &
            \theta \left( e^{-\beta} e^{\pm ix_3} e^{\pm i\Omega_i}; q \right)
            \theta \left( e^{-\beta} e^{\pm ix_4} e^{\pm i\Omega_i}; q \right)
            \theta \left( e^{-\gamma} e^{\pm ix_5} e^{\pm i\Omega_i}; q \right) \Big]
            \label{ladoizquierdoIntriligatorPouliotN=4}
            \numberthis
            \end{align*}
and
            \begin{alignat*}{6}
            \prod_{1 \leq a < b \leq 2N+2} \theta \left( s^2 u_a u_b; q \right) \times
            &=& &
            \Big[ \theta \left( e^{-2\alpha}; q \right) \Big]^2 
            \Big[ \theta \left( e^{-2\beta}; q \right) \Big]^2 
            \theta \left( e^{-2\gamma}; q \right) \\
            & &\times&
            \theta \left( e^{-2\alpha \pm ix_1 \pm ix_2}; q \right)
            \theta \left( e^{-(\alpha + \beta) \pm ix_1 \pm ix_3}; q \right)
            \theta \left( e^{-(\alpha + \beta) \pm ix_1 \pm ix_4}; q \right) \\
            & &\times&
            \theta \left( e^{-(\alpha + \gamma) \pm ix_1 \pm ix_5}; q \right)
            \theta \left( e^{-(\alpha + \beta) \pm ix_2 \pm ix_3}; q \right) \\
            & &\times&
            \theta \left( e^{-(\alpha + \beta) \pm ix_2 \pm ix_4}; q \right)
            \theta \left( e^{-(\alpha + \gamma) \pm ix_2 \pm ix_5}; q \right) \\
            & &\times&
            \theta \left( e^{-2\beta \pm ix_3 \pm ix_4}; q \right)
            \theta \left( e^{-(\beta + \gamma) \pm ix_3 \pm ix_5}; q \right)
            \theta \left( e^{-(\beta + \gamma) \pm ix_4 \pm ix_5}; q \right),
            \label{ladoderechoIntriligatorPouliotN=4}
            \numberthis
            \end{alignat*}
respectively. Again, definitions (\ref{medidaNgeneral}) and (\ref{interactionfactorNgeneral}), crossing parameter (\ref{crossingparameterNgeneral}), and normalization factor $R(\alpha, \beta, \gamma)$ given by
            \begin{align}
            R(\alpha, \beta, \gamma)
            =
            \frac{4! (4 \pi)^4}{(q;q)_{\infty}^{2(4)}}
            \theta \left( e^{2( 2\alpha + 2\beta + \gamma )}; q \right) 
            \Bigg[
            \big[ \theta \left( e^{-2\alpha}; q \right) \big]^2 
            \big[ \theta \left( e^{-2\beta}; q \right)  \big]^2
            \theta \left( e^{-2\gamma}; q \right)
            \Bigg]^{-1},
            \label{normalizationfactorIntriligatorPouliotN=4}
            \end{align}
are considered. Moreover, expressions (\ref{ladoizquierdoIntriligatorPouliotN=4}) and (\ref{ladoderechoIntriligatorPouliotN=4}) lead to write (\ref{indexIntriligatorPouliotnuevoscasosrewritten}) as
            \begin{alignat*}{6}
            \int \left[d\mathbf{\Omega}\right] S \left( \mathbf{\Omega}; q \right)
            \prod_{i=1}^4 \Bigg[
            & \theta \left( e^{-\alpha} e^{\pm ix_1} e^{\pm i\Omega_i}; q \right) 
            \theta \left( e^{-\alpha} e^{\pm ix_2} e^{\pm i\Omega_i}; q \right) \\
            \times &
            \theta \left( e^{-\beta} e^{\pm ix_3} e^{\pm i\Omega_i}; q \right)
            \theta \left( e^{-\beta} e^{\pm ix_4} e^{\pm i\Omega_i}; q \right)
            \theta \left( e^{-\gamma} e^{\pm ix_5} e^{\pm i\Omega_i}; q \right)
            \Bigg]^{-1} \\
            =
            R(\alpha, \beta, \gamma)
            \Bigg[ &
            \theta \left( e^{-2\alpha \pm ix_1 \pm ix_2}; q \right)
            \theta \left( e^{-2\beta \pm ix_3 \pm ix_4}; q \right) \\
            \times &
            \theta \left( e^{-(\eta - \gamma) \pm ix_1 \pm ix_3}; q \right)
            \theta \left( e^{-(\eta - \gamma) \pm ix_1 \pm ix_4}; q \right)
            \theta \left( e^{-(\eta - \gamma) \pm ix_2 \pm ix_3}; q \right) \\
            \times &
            \theta \left( e^{-(\eta - \gamma) \pm ix_2 \pm ix_4}; q \right)
            \theta \left( e^{-(\eta - \beta) \pm ix_1 \pm ix_5}; q \right)
            \theta \left( e^{-(\eta - \beta) \pm ix_2 \pm ix_5}; q \right) \\
            \times &
            \theta \left( e^{-(\eta - \alpha) \pm ix_3 \pm ix_5}; q \right)
            \theta \left( e^{-(\eta - \alpha) \pm ix_4 \pm ix_5}; q \right) \Bigg]^{-1}.
            \numberthis
            \label{indexIntriligatorPouliotnuevoscasosrewrittenN=4}
            \end{alignat*}
            
Finally, definition of the Boltzmann weights as
            \begin{alignat*}{6}
            V_{\alpha} \left( x_1, x_2, \mathbf{\Omega} \right)
            &=& &
                                  \frac{ \displaystyle \prod_{i=1}^4
                                  \big[ \theta \left( e^{-\alpha} e^{\pm ix_1} e^{\pm i\Omega_i}; q \right) 
                                  \theta \left( e^{-\alpha} e^{\pm ix_2} e^{\pm i\Omega_i}; q \right) \big]^{-1}}
                                  { \big[ \theta \left( e^{-2\alpha \pm ix_1 \pm ix_2}; q \right) \big]^{-1}}, \\ \\
            V_{\beta} \left( x_3, x_4, \mathbf{\Omega} \right)
            &=& &
                                  \frac{ \displaystyle \prod_{i=1}^4
                                  \big[ \theta \left( e^{-\beta} e^{\pm ix_3} e^{\pm i\Omega_i}; q \right) 
                                  \theta \left( e^{-\beta} e^{\pm ix_4} e^{\pm i\Omega_i}; q \right) \big]^{-1}}
                                  { \big[ \theta \left( e^{-2\beta \pm ix_3 \pm ix_4}; q \right) \big]^{-1}}, \\
            W_{\gamma} \left( x_5, \mathbf{\Omega} \right)
            &=& &
            \prod_{i=1}^4 \big[ \theta \left( e^{-\gamma} e^{\pm ix_5} e^{\pm i\Omega_i}; q \right) \big]^{-1}, \\ \\
            \overline{V}_{\eta - \alpha} \left( \_, \_, x_3, x_4, x_5 \right)
            &=& 
               \big[& \theta \left( e^{-(\eta - \alpha) \pm ix_3 \pm ix_5}; q \right)
                     \theta \left( e^{-(\eta - \alpha) \pm ix_4 \pm ix_5}; q \right) \big]^{-1}, \\ \\
            \overline{V}_{\eta - \beta} \left( x_1, x_2, \_, \_, x_5 \right)
            &=& 
               \big[& \theta \left( e^{-(\eta - \beta) \pm ix_1 \pm ix_5}; q \right)
                     \theta \left( e^{-(\eta - \beta) \pm ix_2 \pm ix_5}; q \right) \big]^{-1}, \\ \\
            \overline{W}_{\eta - \gamma} \left( x_1, x_2, x_3, x_4, \_ \right)
            &=& 
               \big[& \theta \left( e^{-(\eta - \gamma) \pm ix_1 \pm ix_3}; q \right)
                     \theta \left( e^{-(\eta - \gamma) \pm ix_1 \pm ix_4}; q \right) \\
            & &\times&
                     \theta \left( e^{-(\eta - \gamma) \pm ix_2 \pm ix_3}; q \right)
                     \theta \left( e^{-(\eta - \gamma) \pm ix_2 \pm ix_4}; q \right) \big]^{-1},
            \numberthis
            \label{BoltzmannweightsIntriligatorPouliotN=4}
            \end{alignat*}
allows us to write the index duality (\ref{indexIntriligatorPouliotnuevoscasosrewrittenN=4}) as 
            \begin{align*}
            \int \left[d\mathbf{\Omega}\right] & S \left( \mathbf{\Omega}; q \right)
            V_{\alpha} \left( x_1, x_2, \mathbf{\Omega} \right)
            V_{\beta} \left( x_3, x_4, \mathbf{\Omega} \right)
            W_{\gamma} \left( x_5, \mathbf{\Omega} \right) \\
            = &
            R(\alpha, \beta, \gamma)
            \overline{V}_{\eta - \alpha} \left( \_, \_, x_3, x_4, x_5 \right)
            \overline{V}_{\eta - \beta} \left( x_1, x_2, \_, \_, x_5 \right)
            \overline{W}_{\eta - \gamma} \left( x_1, x_2, x_3, x_4, \_ \right),
            \numberthis
            \label{startriangletyperelationrewrittenIntriligatorN=4}
            \end{align*}
whose analysis is analogous to that of $N=3$ case but with more spin variables.

\subsubsection{Case \texorpdfstring{$N=5$}{}} \label{Case \texorpdfstring{$N=5$}{}}

In this case, expression (\ref{indexIntriligatorPouliot}) can be rewritten as
            \begin{align*}
            \int 
            \frac{(q;q)_{\infty}^{2(5)}}{(5)! (4 \pi)^{5}}
            \left[ \prod_{i=1}^{5} \frac{dz_i}{i z_i} \right] \left[ \prod_{i=1}^{5} \theta \left( z_i^{\pm 2}; q \right)
            \prod_{1 \leq i < j \leq 5} \theta \left( z_i^{\pm 1} z_j^{\pm 1}; q \right)
            \right]
            \prod_{i=1}^{5} \prod_{a=1}^{12} \bigg[ \theta \left( s u_a z_i^{\pm 1}; q \right) \bigg]^{-1} \\
            =
            \theta \left( q s^{-12}; q \right) \prod_{1 \leq a < b \leq 12} \bigg[ \theta \left( s^2 u_a u_b; q \right) \bigg]^{-1}.
            \label{indexIntriligatorPouliotN=5}
            \numberthis
            \end{align*}

A generalization of relations (\ref{expresionesfugacidadIntriligatorPouliot}) is given by
            \begin{align*}
            u_1 &= s^{-1}e^{-\alpha + ix_1},  &  u_3 &= s^{-1}e^{-\alpha + ix_2},  \\
            u_2 &= s^{-1}e^{-\alpha - ix_1},  &  u_4 &= s^{-1}e^{-\alpha - ix_2},  \\
            u_5 &= s^{-1}e^{-\beta + ix_3},  &  u_7 &= s^{-1}e^{-\beta + ix_4},  \\
            u_6 &= s^{-1}e^{-\beta - ix_3},  &  u_8 &= s^{-1}e^{-\beta - ix_4},  \\
            u_9 &= s^{-1}e^{-\gamma + ix_5},  &  u_{11} &= s^{-1}e^{-\gamma + ix_6},  \\
            u_{10} &= s^{-1}e^{-\gamma - ix_5},  &  u_{12} &= s^{-1}e^{-\gamma - ix_6},
            \numberthis
            \label{expresionesfugacidadIntriligatorPouliotN=5}
            \end{align*}
while the new balancing condition reads
            \begin{align}
            \prod_{a=1}^{12} u_a = \frac{1}{q},
            \label{balancingconditionIntriligatorPouliotN=5}
            \end{align}
which in turn implies
            \begin{align}
            q s^{-12} = e^{4(\alpha + \beta + \gamma)}.
            \label{saledebalancingconditionIntriligatorPouliotN=5}
            \end{align}
            
Now, by using Eqs. (\ref{definicionzNgeneral}) and (\ref{expresionesfugacidadIntriligatorPouliotN=5}) for the left and right hand sides of (\ref{indexIntriligatorPouliotN=5}), one has
            \begin{alignat*}{6}
            \prod_{i=1}^5 \prod_{a=1}^{12} \theta \left( s u_a z_i^{\pm 1}; q \right) 
            =
            \prod_{i=1}^5 \bigg[ &
            \theta\left(e^{-\alpha} e^{\pm ix_1} e^{\pm i\Omega_i}; q\right)
            \theta\left(e^{-\alpha} e^{\pm ix_2} e^{\pm i\Omega_i}; q\right) \\
            \times
            &
            \theta\left(e^{-\beta} e^{\pm ix_3} e^{\pm i\Omega_i}; q\right)
            \theta\left(e^{-\beta} e^{\pm ix_4} e^{\pm i\Omega_i}; q\right) \\
            \times
            &
            \theta\left(e^{-\gamma} e^{\pm ix_5} e^{\pm i\Omega_i}; q\right)
            \theta\left(e^{-\gamma} e^{\pm ix_6} e^{\pm i\Omega_i}; q\right) \bigg]
            \numberthis
            \label{ladoizquierdoIntriligatorPouliotN=5}
            \end{alignat*}
and
            \begin{alignat*}{6}
            \prod_{1 \leq a < b \leq 12} \theta \left( s^2 u_a u_b; q \right)
             &=& &
            \bigg[ \theta\left( e^{-2\alpha}; q \right) \theta\left( e^{-2\alpha}; q \right)
                   \theta\left( e^{-2\beta}; q \right) \theta\left( e^{-2\beta}; q \right) 
                   \theta\left( e^{-2\gamma}; q \right) \theta\left( e^{-2\gamma}; q \right) \bigg] \\
            & &\times& 
            \bigg[
            \theta\left( e^{-(\alpha + \alpha)} e^{\pm ix_1} e^{\pm ix_2}; q \right) \\
            & & & \times
            \theta\left( e^{-(\alpha + \beta)} e^{\pm ix_1} e^{\pm ix_3}; q \right)
            \theta\left( e^{-(\alpha + \beta)} e^{\pm ix_1} e^{\pm ix_4}; q \right)  \\
            & & & \times
            \theta\left( e^{-(\alpha + \beta)} e^{\pm ix_2} e^{\pm ix_3}; q \right)
            \theta\left( e^{-(\alpha + \beta)} e^{\pm ix_2} e^{\pm ix_4}; q \right)
            \bigg] \\
            & &\times& 
            \bigg[
            \theta\left( e^{-(\beta + \beta)} e^{\pm ix_3} e^{\pm ix_4}; q \right) \\
            & & & \times
            \theta\left( e^{-(\beta + \gamma)} e^{\pm ix_3} e^{\pm ix_5}; q \right)
            \theta\left( e^{-(\beta + \gamma)} e^{\pm ix_3} e^{\pm ix_6}; q \right) \\
            & & & \times
            \theta\left( e^{-(\beta + \gamma)} e^{\pm ix_4} e^{\pm ix_5}; q \right)
            \theta\left( e^{-(\beta + \gamma)} e^{\pm ix_4} e^{\pm ix_6}; q \right)
            \bigg] \\
            & &\times& 
            \bigg[
            \theta\left( e^{-(\gamma + \gamma)} e^{\pm ix_5} e^{\pm ix_6}; q \right) \\
            & & & \times
            \theta\left( e^{-(\alpha + \gamma)} e^{\pm ix_1} e^{\pm ix_5}; q \right)
            \theta\left( e^{-(\alpha + \gamma)} e^{\pm ix_1} e^{\pm ix_6}; q \right)  \\
            & & & \times
            \theta\left( e^{-(\alpha + \gamma)} e^{\pm ix_2} e^{\pm ix_5}; q \right)
            \theta\left( e^{-(\alpha + \gamma)} e^{\pm ix_2} e^{\pm ix_6}; q \right)
            \bigg],
            \numberthis
            \label{ladoderechoIntriligatorPouliotN=5}
            \end{alignat*}
respectively. Thus, by keeping expressions (\ref{medidaNgeneral}), (\ref{saledebalancingconditionIntriligatorPouliotN=5}), 
(\ref{ladoizquierdoIntriligatorPouliotN=5}) and (\ref{ladoderechoIntriligatorPouliotN=5}) in mind, definition of the interaction factor $S\left(\mathbf{\Omega}; q\right)$ and the normalization factor $R(\alpha, \beta, \gamma)$ as
            \begin{align*}
            S\left(\mathbf{\Omega}; q\right) &=
            \prod_{i=1}^5 \theta \left( e^{\pm 2i\Omega_i}; q \right)
            \prod_{1 \leq i < j \leq 5} \theta \left( e^{\pm i\Omega_i} e^{\pm i\Omega_j}; q \right), \\
            R(\alpha, \beta, \gamma) &= \frac{5! (4 \pi)^5}{(q;q)_{\infty}^{2(5)}}
            \bigg[ \theta \left( e^{4(\alpha + \beta + \gamma)}; q \right) \bigg]
            \bigg[ \theta \left( e^{-2\alpha}; q \right) 
                   \theta \left( e^{-2\beta}; q \right) 
                   \theta \left( e^{-2\gamma}; q \right) \bigg]^{-2},
            \numberthis
            \label{normalizationfactorIntriligatorPouliotN=5}
            \end{align*}
the crossing parameter as (\ref{crossingparameterNgeneral}) and the Boltzmann weights as
            \begin{align*}
            W_{\alpha}^{\pm} \left( x_1, x_2, \mathbf{\Omega} \right)
            &=  
                                             \frac{ \displaystyle
                                             \prod_{i=1}^5 \big[
                                             \theta\left(e^{-\alpha} e^{\pm ix_1} e^{\pm i\Omega_i}; q\right)
                                             \theta\left(e^{-\alpha} e^{\pm ix_2} e^{\pm i\Omega_i}; q\right) \big]^{-1}}
                                             {\big[ \theta\left( e^{-2\alpha} e^{\pm ix_1} e^{\pm ix_2}; q \right) \big]^{-1}}
                                             , \\ \\
            W_{\beta}^{\pm} \left( x_3, x_4, \mathbf{\Omega} \right)
            &=  
                                             \frac{ \displaystyle
                                             \prod_{i=1}^5 \big[
                                             \theta\left(e^{-\beta} e^{\pm ix_3} e^{\pm i\Omega_i}; q\right)
                                             \theta\left(e^{-\beta} e^{\pm ix_4} e^{\pm i\Omega_i}; q\right) \big]^{-1}}
                                             {\big[ \theta\left( e^{-2\beta} e^{\pm ix_3} e^{\pm ix_4}; q \right) \big]^{-1}}
                                             , \\ \\
            W_{\gamma}^{\pm} \left( x_5, x_6, \mathbf{\Omega} \right)
            &=  
                                             \frac{ \displaystyle
                                             \prod_{i=1}^5 \big[
                                             \theta\left(e^{-\gamma} e^{\pm ix_5} e^{\pm i\Omega_i}; q\right)
                                             \theta\left(e^{-\gamma} e^{\pm ix_6} e^{\pm i\Omega_i}; q\right) \big]^{-1}}
                                             {\big[ \theta\left( e^{-2\gamma} e^{\pm ix_5} e^{\pm ix_6}; q \right) \big]^{-1}}
                                             , \\ \\ \\
            W_{\eta-\gamma}(x_1, x_2, x_3, x_4, \_, \_) &= 
            \bigg[
            \theta\left( e^{-(\alpha + \beta)} e^{\pm ix_1} e^{\pm ix_3}; q \right)
            \theta\left( e^{-(\alpha + \beta)} e^{\pm ix_1} e^{\pm ix_4}; q \right)  \\
            & \hspace{0.5cm} \times
            \theta\left( e^{-(\alpha + \beta)} e^{\pm ix_2} e^{\pm ix_3}; q \right)
            \theta\left( e^{-(\alpha + \beta)} e^{\pm ix_2} e^{\pm ix_4}; q \right)
            \bigg]^{-1}, \\
            W_{\eta-\alpha}(\_, \_, x_3, x_4, x_5, x_6) &= 
            \bigg[
            \theta\left( e^{-(\beta + \gamma)} e^{\pm ix_3} e^{\pm ix_5}; q \right)
            \theta\left( e^{-(\beta + \gamma)} e^{\pm ix_3} e^{\pm ix_6}; q \right)  \\
            & \hspace{0.5cm} \times
            \theta\left( e^{-(\beta + \gamma)} e^{\pm ix_4} e^{\pm ix_5}; q \right)
            \theta\left( e^{-(\beta + \gamma)} e^{\pm ix_4} e^{\pm ix_6}; q \right)
            \bigg]^{-1}, \\
            W_{\eta-\beta}(x_1, x_2, \_, \_, x_5, x_6) &= 
            \bigg[
            \theta\left( e^{-(\alpha + \gamma)} e^{\pm ix_1} e^{\pm ix_5}; q \right)
            \theta\left( e^{-(\alpha + \gamma)} e^{\pm ix_1} e^{\pm ix_6}; q \right)  \\
            & \hspace{0.5cm} \times
            \theta\left( e^{-(\alpha + \gamma)} e^{\pm ix_2} e^{\pm ix_5}; q \right)
            \theta\left( e^{-(\alpha + \gamma)} e^{\pm ix_2} e^{\pm ix_6}; q \right)
            \bigg]^{-1},
            \numberthis
            \label{BoltzmannweightsIntriligatorPouliotN=5}
            \end{align*}
makes it possible to rewrite (\ref{indexIntriligatorPouliotN=5}) exactly as the STR type expression
            \begin{align*}
            \int \left[d\mathbf{\Omega}\right] S\left(\mathbf{\Omega}; q\right)
            W_{\alpha}^{\pm} \left( x_1, x_2, \mathbf{\Omega} \right)
            W_{\beta}^{\pm} \left( x_3, x_4, \mathbf{\Omega} \right) 
            W_{\gamma}^{\pm} \left( x_5, x_6, \mathbf{\Omega} \right) \hspace{5cm} \\
            =
            R(\alpha, \beta, \gamma)
            W_{\eta-\alpha}(\_, \_, x_3, x_4, x_5, x_6)
            W_{\eta-\beta}(x_1, x_2, \_, \_, x_5, x_6)
            W_{\eta-\gamma}(x_1, x_2, x_3, x_4, \_, \_),
            \numberthis
            \label{startriangletyperelationrewrittenIntriligatorN=5}
            \end{align*}
which can be thought of as a generalization of $N=2$ case for more spin variables.
            
\subsubsection{General case \texorpdfstring{$N=3k+2$}{}} \label{Case \texorpdfstring{$N=3k+2$}{}}

Calculations made for the cases $N=2$ and $N=5$ can be naturally extended to all values of $N$ such that $N=3k+2$ for any 
$k \in \mathbb{Z}^+ \cup \{ 0 \}$\footnote{The cases $k=0$ and $k=1$ correspond precisely to the cases $N=2$ and $N=5$, respectively.}. In this case, (\ref{indexIntriligatorPouliot}) can be rewritten as
            \begin{align*}
            \int R(k)
            \left[ \prod_{i=1}^{3k+2} \frac{dz_i}{i z_i} \right] \left[ \prod_{i=1}^{3k+2} \theta \left( z_i^{\pm 2}; q \right)
            \prod_{1 \leq i < j \leq 3k+2} \theta \left( z_i^{\pm 1} z_j^{\pm 1}; q \right)
            \right]
            \prod_{i=1}^{3k+2} \prod_{a=1}^{6(k+1)} \bigg[ \theta \left( s u_a z_i^{\pm 1}; q \right) \bigg]^{-1} \\
            =
            \theta \left( q s^{-6(k+1)}; q \right) \prod_{1 \leq a < b \leq 6(k+1)} \bigg[ \theta \left( s^2 u_a u_b; q \right) \bigg]^{-1},
            \label{indexIntriligatorPouliotN=3k+2}
            \numberthis
            \end{align*}
where
            \begin{align}
            R(k) = \frac{(q;q)_{\infty}^{2(3k+2)}}{(3k+2)! (4 \pi)^{(3k+2)}}.
            \label{normalizationR(k)}
            \end{align}

Generalization of relations (\ref{expresionesfugacidadIntriligatorPouliot}) is
            \begin{align*}
            u_1 &= s^{-1}e^{-\alpha + ix_1},  &  &\cdots  &  u_{2(k+1)-1} &= s^{-1}e^{-\alpha + ix_{(k+1)}},  \\
            u_2 &= s^{-1}e^{-\alpha - ix_1},  &  &\cdots  &  u_{2(k+1)}   &= s^{-1}e^{-\alpha - ix_{(k+1)}},  \\ \\
            u_{2(k+1)+1} &= s^{-1}e^{-\beta + ix_{(k+1)+1}},  &  &\cdots  &  u_{4(k+1)-1} &= s^{-1}e^{-\beta + ix_{2(k+1)}},  \\
            u_{2(k+1)+2} &= s^{-1}e^{-\beta - ix_{(k+1)+1}},  &  &\cdots  &  u_{4(k+1)} &= s^{-1}e^{-\beta - ix_{2(k+1)}},  \\ \\
            u_{4(k+1)+1} &= s^{-1}e^{-\gamma + ix_{2(k+1)+1}},  &  &\cdots  &  u_{6(k+1)-1} &= s^{-1}e^{-\gamma + ix_{3(k+1)}},  \\
            u_{4(k+1)+2} &= s^{-1}e^{-\gamma - ix_{2(k+1)+1}},  &  &\cdots  &  u_{6(k+1)} &= s^{-1}e^{-\gamma - ix_{3(k+1)}},
            \numberthis
            \label{expresionesfugacidadIntriligatorPouliotN=3k+2}
            \end{align*}
while the new balancing condition reads
            \begin{align}
            \prod_{a=1}^{6(k+1)} u_a = \frac{1}{q}.
            \label{balancingconditionIntriligatorPouliotN=3k+2}
            \end{align}
           
Then, by keeping crossing parameter (\ref{crossingparameterNgeneral}) and expressions (\ref{medidaNgeneral}), (\ref{expresionesfugacidadIntriligatorPouliotN=3k+2}) and (\ref{balancingconditionIntriligatorPouliotN=3k+2}) in mind, generalization of interaction and normalization factors, $S\left(\mathbf{\Omega}; q\right)$ and $R(\alpha, \beta, \gamma)$, respectively, to
            \begin{align*}
            S\left(\mathbf{\Omega}; q\right) &=
            \prod_{i=1}^{3k+2} \theta \left( e^{\pm 2i\Omega_i}; q \right)
            \prod_{1 \leq i < j \leq 3k+2} \theta \left( e^{\pm i\Omega_i} e^{\pm i\Omega_j}; q \right), \\
            R(\alpha, \beta, \gamma) &= \frac{(3k+2)! (4 \pi)^{3k+2}}{(q;q)_{\infty}^{2(3k+2)}}
            \bigg[ \theta \left( e^{2(k+1) (\alpha + \beta + \gamma)}; q \right) \bigg]
            \bigg[ \theta \left( e^{-2\alpha}; q \right) 
                   \theta \left( e^{-2\beta}; q \right) 
                   \theta \left( e^{-2\gamma}; q \right) \bigg]^{-(k+1)},
            \numberthis
            \label{normalizationfactorIntriligatorPouliotN=3k+2}
            \end{align*}
and Boltzmann weights to
            \begin{alignat*}{6}
            W_{\alpha}^{\pm} \left( x_1, \dots, x_{k+1}, \mathbf{\Omega} \right) 
            &=& & 
            \frac{ \displaystyle 
            \prod_{i=1}^{3k+2} \prod_{j=1}^{k+1} \bigg[
            \theta\left(e^{-\alpha} e^{\pm ix_j} e^{\pm i\Omega_i}; q\right) \bigg]^{-1}}
            {\displaystyle \prod_{\substack{m<n \\ m,n=1, \dots, (k+1)}}
            \bigg[ \theta\left( e^{-2\alpha} e^{\pm ix_m} e^{\pm ix_n}; q \right) \bigg]^{-1}}
            , \\ \\
            W_{\beta}^{\pm} \left( x_{(k+1)+1}, \dots, x_{2(k+1)}, \mathbf{\Omega} \right) 
            &=& &
            \frac{ \displaystyle 
            \prod_{i=1}^{3k+2} \prod_{j=(k+1)+1}^{2(k+1)} \bigg[
            \theta\left(e^{-\beta} e^{\pm ix_j} e^{\pm i\Omega_i}; q\right) \bigg]^{-1}}
            {\displaystyle \prod_{\substack{m<n \\ m,n=(k+1)+1, \dots, 2(k+1)}}
            \bigg[ \theta\left( e^{-2\beta} e^{\pm ix_m} e^{\pm ix_n}; q \right) \bigg]^{-1}}
            , \\ \\
            W_{\gamma}^{\pm} \left( x_{2(k+1)+1}, \dots, x_{3(k+1)}, \mathbf{\Omega} \right) 
            &=& &
            \frac{ \displaystyle
            \prod_{i=1}^{3k+2} \prod_{j=2(k+1)+1}^{3(k+1)}
                                             \bigg[ \theta\left(e^{-\gamma} e^{\pm ix_j} e^{\pm i\Omega_i}; q\right) \bigg]^{-1}}
            {\displaystyle \prod_{\substack{m<n \\ m,n=2(k+1)+1, \dots, 3(k+1)}}
            \bigg[ \theta\left( e^{-2\gamma} e^{\pm ix_m} e^{\pm ix_n}; q \right) \bigg]^{-1}}
            , \\
            \end{alignat*}
            \begin{alignat*}{6}
            W_{\eta-\gamma} \left( \{x_l\}_{l=1, \dots, 3(k+1)} \backslash \{x_{2(k+1)+1, \dots, 3(k+1)}\} \right)
            &=& & 
            \prod_{\substack{m=1, \dots, (k+1) \\ n=(k+1)+1, \dots, 2(k+1)}} 
            \bigg[ \theta\left( e^{-(\alpha + \beta)} e^{\pm ix_m} e^{\pm ix_n}; q \right) \bigg]^{-1}, \\ \\
            W_{\eta-\alpha} \left( \{x_l\}_{l=1, \dots, 3(k+1)} \backslash \{x_1, \dots, x_{(k+1)}\} \right) 
            &=& & 
            \prod_{\substack{m=(k+1)+1, \dots, 2(k+1) \\ n=2(k+1)+1, \dots, 3(k+1)}} 
            \bigg[ \theta\left( e^{-(\beta + \gamma)} e^{\pm ix_m} e^{\pm ix_n}; q \right) \bigg]^{-1}, \\ \\
            W_{\eta-\beta} \left( \{x_l\}_{l=1, \dots, 3(k+1)} \backslash \{x_{(k+1)+1, \dots, 2(k+1)}\} \right) 
            &=& & 
            \prod_{\substack{m=1, \dots, (k+1) \\ n=2(k+1)+1, \dots, 3(k+1)}} 
            \bigg[ \theta\left( e^{-(\alpha + \gamma)} e^{\pm ix_m} e^{\pm ix_n}; q \right) \bigg]^{-1},
            \numberthis
            \label{BoltzmannweightsIntriligatorPouliotN=3k+2}
            \end{alignat*}
leads us to write down the index duality (\ref{indexIntriligatorPouliotN=3k+2}) as the STR type expression
            \begin{align*}
            \int \left[d\mathbf{\Omega}\right] S\left(\mathbf{\Omega}; q\right)
            W_{\alpha}^{\pm} \left( x_1, \dots, x_{k+1}, \mathbf{\Omega} \right)
            W_{\beta}^{\pm} \left( x_{(k+1)+1}, \dots, x_{2(k+1)}, \mathbf{\Omega} \right)
            W_{\gamma}^{\pm} \left( x_{2(k+1)+1}, \dots, x_{3(k+1)}, \mathbf{\Omega} \right)
            & \\
            =
            R(\alpha, \beta, \gamma)
            W_{\eta-\alpha} \left( \{x_l\}_{l=1, \dots, 3(k+1)} \backslash \{x_1, \dots, x_{(k+1)}\} \right)
            W_{\eta-\beta} \left( \{x_l\}_{l=1, \dots, 3(k+1)} \backslash \{x_{(k+1)+1, \dots, 2(k+1)}\} \right)
            & \\ \times 
            W_{\eta-\gamma} \left( \{x_l\}_{l=1, \dots, 3(k+1)} \backslash \{x_{2(k+1)+1, \dots, 3(k+1)}\} \right) & ,
            \numberthis
            \label{startriangletyperelationrewrittenIntriligatorN=3k+2}
            \end{align*}
which is a generalization of the results obtained in subsections \ref{Case \texorpdfstring{$N=2$}{}} and \ref{Case \texorpdfstring{$N=5$}{}}.
            
\subsection{\texorpdfstring{$2d$ $\mathcal{N}=(0,2)$ $USp(2N)$}{} Csáki-Skiba-Schmaltz duality} \label{CsakiSkibaSchmaltzduality}

In this subsection we study a $2d$ $\mathcal{N}=(0,2)$ $USp(2N)$ Csáki-Skiba-Schmaltz duality for theories with matter in the antisymmetric tensor representation. This duality is obtained in \cite{Sacchi:2020pet}\footnote{This reference actually found two different index dualities, we refer here to the one given by $N_b=4$ and $N_f=0$.}
from dimensional reduction of $4d$ $\mathcal{N}=1$ $USp(2N)$ Csáki-Skiba-Schmaltz duality for theories with matter in the antisymmetric tensor representation first studied in \cite{Csaki:1996eu}. The $2d$ $\mathcal{N}=(0,2)$ $USp(2N)$ Csáki-Skiba-Schmaltz duality is given between a $USp(2N)$ gauge theory with $4$ chiral multiplets in the fundamental representation, $N$ Fermi multiplets and one antisymmetric chiral, and a Laudau-Ginzburg model with $6N$ chiral multiplets and $N$ Fermi multiplets. The elliptic flavoured genera expression for the duality of these $2d$ $\mathcal{N}=(0,2)$ $USp(2N)$ supersymmetric quiver gauge theories is, from \cite{Sacchi:2020pet},
            \begin{align*}
            \prod_{i=1}^{N} \theta\left( qx^{-i};q \right)
            \int
            \left[
            \frac{d \bar{z}_N}{[\theta(x; q)]^N \prod_{1 \leq i < j \leq N} \theta\left( x z_i^{\pm 1} z_j^{\pm 1}; q \right)}
            \right]
            \left[
            \frac{1}{\prod_{i=1}^N \prod_{a=1}^{4} \theta \left( s x^{\frac{1-N}{3}} u_a z_i^{\pm 1}; q \right)}
            \right] \\
            =
            \prod_{i=1}^N
            \frac{\theta \left( q s^{-4} x^{i-\frac{2N+1}{3}}; q \right)}
                 {\prod_{1 \leq a < b \leq 4} \theta \left( s^2 x^{i-\frac{2N+1}{3}} u_a u_b; q \right)},
            \numberthis
            \label{indexCsakiSkibaSchmaltz}
            \end{align*}
where, again,
            \begin{align}
            d \bar{z}_N = \frac{(q;q)_{\infty}^{2N}}{N! (4 \pi)^N} 
            \prod_{i=1}^N \left[ \frac{dz_i}{i z_i} \theta \left( z_i^{\pm 2}; q \right) \right]
            \prod_{1 \leq i < j \leq N} \theta \left( z_i^{\pm 1} z_j^{\pm 1}; q \right)
            \label{medidaCsakiSkibaSchmaltz}
            \end{align}
is the measure associated with $USp(2N)$. Here, $\{u_a\}_{a=1, \dots, 4}$, $\{s\}$ and $\{x\}$ are the sets of fugacities associated with the global symmetry group $SU(4)_u \times U(1)_s \times U(1)_x$ of the theories. Note that for $N=1$ the duality is, as well as in the Intriligator-Pouliot case of section \ref{Case \texorpdfstring{$N=1$}{}}, reduced to the $2d$ $\mathcal{N}=(0,2)$ $SU(2)$ duality considered in \cite{Dedushenko:2017osi,Gadde:2015wta}.

Index duality (\ref{indexCsakiSkibaSchmaltz}) can be analysed for general $N$ by defining, in analogy with (\ref{expresionesfugacidadIntriligatorPouliotN=1}), the following relations between fugacities, spectral parameters and spin variables
            \begin{align*}
            u_1 &= s^{-1} x^{\frac{N-1}{3}} e^{-\alpha + ix_1},  &  u_3 &= s^{-1} x^{\frac{N-1}{3}} e^{-\beta + ix_2}, \\
            u_2 &= s^{-1} x^{\frac{N-1}{3}} e^{-\alpha - ix_1},  &  u_4 &= s^{-1} x^{\frac{N-1}{3}} e^{-\beta - ix_2},
            \numberthis
            \label{expresionesfugacidadCsakiSkibaSchmaltz}
            \end{align*}
and the balancing condition
            \begin{align}
            \prod_{a=1}^4 u_a = \frac{1}{q}.
            \label{balancingconditionCsakiSkibaSchmaltz}
            \end{align}
            
First of all, use (\ref{definicionzNgeneral}) to define
            \begin{align}
            S'\left( \mathbf{\Omega}; x, q\right) 
            = 
            \prod_{i=1}^{N} \theta\left( qx^{-i};q \right)
            \left[
            \frac{\displaystyle \prod_{i=1}^N \theta \left( e^{\pm 2i\Omega_i}; q \right)
                  \prod_{1 \leq i < j \leq N} \theta \left( e^{\pm i\Omega_i} e^{\pm i\Omega_j}; q \right)}
                 {\displaystyle [\theta(x; q)]^N \prod_{1 \leq i < j \leq N} \theta\left( x e^{\pm i\Omega_i} e^{\pm i\Omega_j}; q \right)}
            \right]
            \label{interactiontermauxiliarCsakiSkibaSchmaltz}
            \end{align}
and, in turn, use it to rewrite (\ref{indexCsakiSkibaSchmaltz}) as
            \begin{align*}
            \frac{(q;q)_{\infty}^{2N}}{N! (4 \pi)^N}
            \int
            \left[ \prod_{i=1}^N \frac{dz_i}{i z_i} \right]
            \bigg[ S'\left( \mathbf{\Omega}; x, q\right) \bigg]
            \left[
            \frac{1}{\prod_{i=1}^N \prod_{a=1}^{4} \theta \left( s x^{\frac{1-N}{3}} u_a z_i^{\pm 1}; q \right)}
            \right] \\
            =
            \prod_{i=1}^N
            \frac{\theta \left( q s^{-4} x^{i-\frac{2N+1}{3}}; q \right)}
                 {\prod_{1 \leq a < b \leq 4} \theta \left( s^2 x^{i-\frac{2N+1}{3}} u_a u_b; q \right)}.
            \numberthis
            \label{indexCsakiSkibaSchmaltzrewritten}
            \end{align*}
            
By using expressions (\ref{expresionesfugacidadCsakiSkibaSchmaltz}) it is possible to rewrite some factors in (\ref{indexCsakiSkibaSchmaltzrewritten}) as
            \begin{align*}
            \prod_{i=1}^N \prod_{a=1}^{4} \theta \left( s x^{\frac{1-N}{3}} u_a z_i^{\pm 1}; q \right) 
            =
            \prod_{i=1}^N \bigg[
                   & \theta \left( e^{-\alpha + ix_1} e^{\pm i\Omega_i}; q \right) \theta \left( e^{-\alpha - ix_1} e^{\pm i\Omega_i}; q \right) 
                     \theta \left( e^{-\beta + ix_2} e^{\pm i\Omega_i}; q \right)  \\
            \times & \theta \left( e^{-\beta - ix_2} e^{\pm i\Omega_i}; q \right) \bigg] \\
            =
            \prod_{i=1}^N \bigg[
            & \theta \left( e^{-\alpha} e^{\pm ix_1} e^{\pm i\Omega_i}; q \right) 
              \theta \left( e^{-\beta} e^{\pm ix_2} e^{\pm i\Omega_i}; q \right) \bigg]
            \label{ladoizquierdoCsakiSkibaSchmaltz}
            \numberthis
            \end{align*}
and
            \begin{align*}
            \prod_{i=1}^N \prod_{1 \leq a < b \leq 4} \theta \left( s^2 x^{i-\frac{2N+1}{3}} u_a u_b; q \right)
            =
            \prod_{i=1}^N \bigg[
            & \theta \left( x^{i-1} e^{-2\alpha}; q \right) \theta \left( x^{i-1} e^{-2\beta}; q \right)  \\
            \times
            & \theta \left( x^{i-1} e^{-(\alpha + \beta) + ix_1 + ix_2}; q \right) \theta \left( x^{i-1} e^{-(\alpha + \beta) + ix_1 - ix_2}; q \right)  \\
            \times
            & \theta \left( x^{i-1} e^{-(\alpha + \beta) - ix_1 + ix_2}; q \right) \theta \left( x^{i-1} e^{-(\alpha + \beta) - ix_1 - ix_2}; q \right) \bigg] \\
            =
            \prod_{i=1}^N \bigg[
            & \theta \left( x^{i-1} e^{-2\alpha}; q \right) \theta \left( x^{i-1} e^{-2\beta}; q \right) 
            \theta \left( x^{i-1} e^{-(\alpha + \beta)} e^{\pm ix_1} e^{\pm ix_2}; q \right) \bigg].
            \label{ladoderechoCsakiSkibaSchmaltz}
            \numberthis
            \end{align*}

Also, we note that the balancing condition (\ref{balancingconditionCsakiSkibaSchmaltz}) implies the relation
            \begin{align}
            qs^{-4} x^{\frac{4(N-1)}{3}} = e^{2(\alpha + \beta)},
            \label{saledebalancingconditionCsakiSkibaSchmaltz}
            \end{align}
from where we have
            \begin{align}
            \theta \left( q s^{-4} x^{i-\frac{2N+1}{3}}; q \right) = \theta \left( x^{i-(2N-1)} e^{2(\alpha + \beta)}; q \right).
            \label{saledebalancingconditionCsakiSkibaSchmaltz2}
            \end{align}            
            
By using expressions (\ref{ladoizquierdoCsakiSkibaSchmaltz}), (\ref{ladoderechoCsakiSkibaSchmaltz}) and (\ref{saledebalancingconditionCsakiSkibaSchmaltz2}), the index duality (\ref{indexCsakiSkibaSchmaltzrewritten}) rewrites as
            \begin{align*}
            \frac{(q;q)_{\infty}^{2N}}{N! (4 \pi)^N}
            \int
            \left[ \prod_{i=1}^N d\Omega_i \right]
            \bigg[ S'\left( \mathbf{\Omega}; x, q\right) \bigg]
            \prod_{i=1}^N
            \bigg[
            \theta \left( e^{-\alpha} e^{\pm ix_1} e^{\pm i\Omega_i}; q \right) 
            \theta \left( e^{-\beta} e^{\pm ix_2} e^{\pm i\Omega_i}; q \right)
            \bigg]^{-1} \\
            =
            \theta \left( x^{i-(2N-1)} e^{2(\alpha + \beta)}; q \right)
            \prod_{i=1}^N
            \bigg[
            \theta \left( x^{i-1} e^{-2\alpha}; q \right) \theta \left( x^{i-1} e^{-2\beta}; q \right) 
            \theta \left( x^{i-1} e^{-(\alpha + \beta)} e^{\pm ix_1} e^{\pm ix_2}; q \right) \bigg]^{-1}.
            \numberthis
            \label{indexCsakiSkibaSchmaltzrewritten2}
            \end{align*}
            
Thus, by keeping (\ref{medidaNgeneral}) in mind, identification of the interaction factor $S\left(\mathbf{\Omega}; x, q\right)$ and the normalization factor $R(\alpha, \beta; x)$ as
            \begin{align*}
            S\left(\mathbf{\Omega}; x, q\right) &= S'\left( \mathbf{\Omega}; x, q\right), \\
            R(\alpha, \beta; x) &= \frac{N! (4 \pi)^N}{(q;q)_{\infty}^{2N}}
            \bigg[ \theta \left( x^{i-(2N-1)} e^{2(\alpha + \beta)}; q \right) \bigg]
            \prod_{i=1}^N
            \bigg[
            \theta \left( x^{i-1} e^{-2\alpha}; q \right) \theta \left( x^{i-1} e^{-2\beta}; q \right) 
            \bigg]^{-1},
            \numberthis
            \label{normalizationfactorCsakiSkibaSchmaltz}
            \end{align*}
and the Boltzmann weights as
            \begin{align*}
            W_{\alpha} \left( x_1, \mathbf{\Omega} \right) &= \prod_{i=1}^N
            \bigg[ \theta \left( e^{-\alpha} e^{\pm ix_1} e^{\pm i\Omega_i}; q \right) \bigg]^{-1}, \\
            W_{\beta} \left( x_2, \mathbf{\Omega} \right) &= \prod_{i=1}^N
            \bigg[ \theta \left( e^{-\beta} e^{\pm ix_2} e^{\pm i\Omega_i}; q \right) \bigg]^{-1}, \\
            W_{\alpha + \beta}^x \left( x_1, x_2 \right) &= \prod_{i=1}^N
            \bigg[ \theta \left( x^{i-1} e^{-(\alpha + \beta)} e^{\pm ix_1} e^{\pm ix_2}; q \right) \bigg]^{-1},
            \numberthis
            \label{BoltzmannweightsCsakiSkibaSchmaltz}
            \end{align*}
leads us to put the index duality (\ref{indexCsakiSkibaSchmaltzrewritten2}) in the form
            \begin{align}
            \int \left[ d\mathbf{\Omega} \right] S\left(\mathbf{\Omega}; x, q\right)
            W_{\alpha} \left( x_1, \mathbf{\Omega} \right)
            W_{\beta} \left( x_2, \mathbf{\Omega} \right)
            =
            R(\alpha, \beta; x) 
            W_{\alpha + \beta}^x \left( x_1, x_2 \right).
            \label{startriangletyperelationrewrittenCsakiSkibaSchmaltz}
            \end{align}
            
Again, as in the Intriligator-Pouliot case discussed in section \ref{Case \texorpdfstring{$N=1$}{}}, expression (\ref{startriangletyperelationrewrittenCsakiSkibaSchmaltz}) has an analogous form to that of the triangle identity considered in \cite{Kels:2018xge} but with a slight distinction between left and right hand side Boltzmann weights given by an extra parameter $x$ in the latter one, this situation is quite similar to that in expression (\ref{BoltzmannweightsSpiridonov}) for Boltzmann weights found in \cite{Spiridonov:2010em}.
            
\section{Final Remarks}

In the present paper, a brief overview of the gauge/YBE correspondence is provided. The work performed in this subject is brand new, the dictionary of this correspondence is incomplete and it is yet under construction.  For example, references \cite{Spiridonov:2009za,Spiridonov:2011hf} contain a large list of $4d$ $\mathcal{N}=1$ dualities and their corresponding supersymmetric index equalities, this is done for many gauge groups, but it is not clear if there are star-triangle type relations associated to all of them.
In particular, star-triangle type relations associated to $2d$ ${\cal N}=(0,2)$ supersymmetric quiver gauge theory dualities had not been found before and so they are not included yet into this context. It was the purpose of this article to provide them for the $2d$ ${\cal N}=(0,2)$ $USp(2N)$ dualities presented in Ref. \cite{Sacchi:2020pet}, that is, Intriligator-Pouliot and Csáki-Skiba-Schmaltz dualities in two dimensions.

The derivation coming from Intriligator-Pouliot duality for $\mathcal{N}=(0,2)$ supersymmetric quiver gauge theories was carried out explicitly for different values of $N$.
In particular, for $N=2,5$ we found that the realization of the duality conditions (\ref{indexIntriligatorPouliotN=2}) and (\ref{indexIntriligatorPouliotN=5}) implied the corresponding STR type expressions (\ref{startriangletyperelationrewrittenIntriligator}) and (\ref{startriangletyperelationrewrittenIntriligatorN=5}), respectively. These cases were generalized to the value $N=3k+2$, where duality condition (\ref{indexIntriligatorPouliotN=3k+2}) implied STR type expression (\ref{startriangletyperelationrewrittenIntriligatorN=3k+2}). All these STR type expressions have two different definitions for Boltzmann weights, one for the left hand side and other for the right hand side ones.
The cases with $N=3,4$ (expressions (\ref{startriangletyperelationrewrittenIntriligatorN=3}) and (\ref{startriangletyperelationrewrittenIntriligatorN=4}), respectively) are somewhat similar to the asymmetric form of the star-triangle relation already reported in Refs. \cite{Kels:2019ktt,Kels:2018xge,Kels:2020zjn} although they are not exactly the same.
The value $N=1$, our expression (\ref{startriangletyperelationrewrittenIntriligatorN=1}), is more interesting because it highly resembles the triangle identity reported previously \cite{Kels:2018xge} in the literature of Yang–Baxter/$3D$-consistency correspondence.

The derivation of STR type expression (\ref{startriangletyperelationrewrittenCsakiSkibaSchmaltz}) from Csáki-Skiba-Schmaltz duality for $\mathcal{N}=(0,2)$ supersymmetric quiver gauge theories with an antisymmetric tensor is valid for all values of $N$, this expression also resembles the triangle identity found in \cite{Kels:2018xge} but the right hand side Boltzmann weight have an extra parameter similar in spirit to definitions (\ref{BoltzmannweightsSpiridonov}) coming from \cite{Spiridonov:2010em}. 

As part of the work, the Boltzmann weights as well as the interaction and normalization factors were completely determined for all cases. It is worthy to remark that all examples we found here have not exactly the form of a SSR or a STR expression, which we certainly know give rise to integrable models. Thus, although we have shown that the diverse dualities of certain $2d$ $\mathcal{N}=(0,2)$ models considered here have an associated STR type expression they probably do not represent integrable models. Actually, in the context of Yang–Baxter/$3D$-consistency correspondence the relation of triangle identity with integrability is still unclear \cite{Kels:2018xge}.
We hope our expressions could give insights in the study of integrability in an alternative direction to that of Ref. \cite{Gadde:2013lxa} were a triality between $2d$ ${\cal N}=(0,2)$ models was found and it was conjectured that a tetrahedron equation of certain integrable systems would be associated with those models. Precise determination of integrability properties of our STR type expressions is an interesting topic for future work.

We conclude the paper by mentioning that it would be interesting to analyse the relation of star-triangle type relations, no matter whether they have associated integrable models or not, and topological knot invariants (see, for instance, \cite{Spiridonov:2011hf,Wu:1992zzb}). Moreover, it would also be interesting to find a description of the gauge/YBE correspondence in terms of brane box configurations discussed in Refs. \cite{GarciaCompean:1998kh,Franco:2017cjj}.

\vskip 2truecm
\centerline{\bf Acknowledgements}

The work of J. de-la-Cruz-Moreno was supported by a CONACyT graduate fellowship. 

\vskip 2truecm
	    
\end{document}